# OPERATIONAL AGENCY: A PERMEABLE LEGAL FICTION FOR TRACING CULPABILITY IN AI SYSTEMS

Anirban Mukherjee*

Hannah Hanwen Chang†

* Anirban Mukherjee (anirban@avyayamholdings.com) is Principal at Avyayam Holdings.

†Hannah H. Chang (hannahchang@smu.edu.sg; corresponding author) is Associate Professor of Marketing at the Lee Kong Chian School of Business, Singapore Management University.

This research was supported by the Ministry of Education (MOE), Singapore, under its Academic Research Fund (AcRF) Tier 2 Grant, No. MOE-T2EP40124-0005.

**Abstract**

Modern artificial intelligence (AI) systems act with a high degree of independence yet lack legal personhood—a paradox that fractures doctrines grounded in human-centric notions of *mens rea* and *actus reus*. This Article introduces *Operational Agency (OA)*—a permeable legal fiction structured as an *ex post* evidentiary framework—and *Operational Agency Graph (OAG)*, a tool for mapping causal interactions among human actors, organizations, and AI systems. OA evaluates an AI's observable operational characteristics: its *goal-directedness* (as a proxy for intent), *predictive processing* (as a proxy for foresight), and *safety architecture* (as a proxy for a standard of care). OAG operationalizes that analysis by embedding these characteristics in a causal graph to trace and apportion culpability among developers, fine-tuners, deployers, and users. Drawing on corporate criminal liability, the innocent-agent doctrine, and secondary and vicarious liability frameworks, the Article shows how OA and OAG strengthen existing doctrines. Across five real-world case studies spanning tort, civil rights, constitutional law, and antitrust, it demonstrates how the framework addresses challenges ranging from autonomous vehicle collisions to algorithmic price-fixing, offering courts a principled evidentiary method—and legislatures and industry a conceptual foundation—to ensure human accountability keeps pace with technological autonomy, without conferring personhood on AI.

---

TABLE OF CONTENTS

## INTRODUCTION

The law has long crafted powerful yet pragmatic legal fictions to bridge gaps between operational capacity and legal accountability. Under Roman law, slaves lacked legal personhood.[1] Yet, the *peculium* system held masters liable for their slaves' commercial transactions, calibrated to the extent of authorization.[2] Modern law developed *corporate*

---

[1] *See* ALAN WATSON, ROMAN SLAVE LAW 46-66 (1987); W.W. BUCKLAND, THE ROMAN LAW OF SLAVERY 1-9 (1908).

[2] Slaves granted a *peculium* (a fund or property for management) could transact and conduct business, with creditors given praetorian actions against the master: Full liability under the *actio quod iussu* (action for acting on orders) if the slave acted on the master's direct orders, or liability limited to the value of the *peculium* under the *actio de peculio* (action concerning the peculium) if the transaction related to its management. *See* Watson, *supra* note 1, at 90-101; Buckland, *supra* note 1, at 187-238.

*personhood* to hold firms accountable.[3] Corporations could sue and be sued, enter contracts, and bear civil liability, even as this fiction strained criminal law concepts designed for natural persons—*mens rea*, moral culpability, imprisonment.[4] Courts responded with doctrines like *respondeat superior* (holding corporations vicariously liable for employees' acts within the scope of employment) and senior-officer attribution (imputing the intent of high-level officials to the corporation itself), but debates persist over whether monetary penalties can adequately punish—or deter—entities without bodies to incarcerate or souls to condemn.[5]

Today, Artificial Intelligence (AI) presents the latest iteration of this paradox. Modern AI systems pursue long-term goals, make decisions, and execute complex workflows without continuous human oversight.[6] Unlike traditional generative AI systems, which are limited to executing well-defined tasks (e.g., "generate an image"), these systems proactively orchestrate multi-step processes to pursue open-ended objectives, such as autonomously searching the web, synthesizing findings, writing code, and deploying another AI to complete a subtask.[7] Yet, this

---

[3] *See* Trs. of Dartmouth Coll. v. Woodward, 17 U.S. (4 Wheat.) 518, 636 (1819) (describing the corporation as "an artificial being, invisible, intangible, and existing only in contemplation of law").

[4] *See* N.Y. Cent. & Hudson River R.R. Co. v. United States, 212 U.S. 481, 492-93 (1909) (upholding corporate criminal liability and rejecting the doctrine that corporations could not commit crimes).

[5] *See N.Y. Cent.,* 212 U.S. at 494–96 (extending respondeat superior to corporate criminal liability); John C. Coffee, Jr., *"No Soul to Damn: No Body to Kick": An Unscandalized Inquiry into the Problem of Corporate Punishment*, 79 MICH. L. REV. 386 (1981). *See also* Jennifer Arlen, *Corporate Criminal Liability: Theory and Evidence*, *in* RESEARCH HANDBOOK ON THE ECONOMICS OF CRIMINAL LAW 144 (Alon Harel & Keith N. Hylton eds., 2012) (analyzing the economic structure of corporate criminal liability and arguing that optimal enforcement requires duty-based liability calibrated to a firm's investment in monitoring and self-policing).

[6] *See* Alan Chan, Carson Ezell, Max Kaufmann, Kevin Wei, Lewis Hammond, Herbie Bradley, Emma Bluemke, Nitarshan Rajkumar, David Krueger, Noam Kolt, Lennart Heim, Markus Anderljung, *Visibility into AI Agents*, in Proceedings of the 2024 ACM Conference on Fairness, Accountability, and Transparency (FAccT '24) 958, 958-59 (2024), https://dl.acm.org/doi/epdf/10.1145/3630106.3658948 (defining AI agents as systems pursuing complex goals with limited supervision that may take multiple consequential actions before a human intervenes).

[7] In prior work, we term these capabilities *fluid agency* for their (i) *stochastic*, (ii) *dynamic*, and (iii) *adaptive* nature. Fluid agency irreducibly entangles human and machine contributions and control, creating a fundamental *unmappability* that fractures traditional frameworks. That Article proposes functional equivalence as a stabilizing principle and identifies, among alternative liability approaches, a "doctrinally conservative path": developing evidentiary tools that analyze an AI's operational characteristics as a proxy for human fault. This Article provides the

leap in technical capacity has not been matched by a corresponding evolution in legal status.[8] Under current law, AI remains mere property, lacking legal personhood (i.e., the capacity to be a rights-holder or duty-bearer).[9] It holds no rights, bears no duties, and therefore incurs no direct liability.

Herein lies the rupture. Tort assumes a clear line of sight to either a low-autonomy tool controlled by a legal person, or a high-autonomy person whose own legal status grounds

---

systematic framework for that approach. *See* Anirban Mukherjee & Hannah H. Chang, *Fluid Agency in AI Systems: A Case for Functional Equivalence in Copyright, Patent, and Tort*, 21 WASH. J.L. TECH. & ARTS 1 (2026), https://digitalcommons.law.uw.edu/wjlta/vol21/iss1/3 [https://perma.cc/4LNK-FLSG] [hereinafter *Fluid Agency*]. *See also* Yonadav Shavit, Sandhini Agarwal, Miles Brundage, Steven Adler, Cullen O'Keefe, Rosie Campbell, Teddy Lee, Pamela Mishkin, Tyna Eloundou, Alan Hickey, Katarina Slama, Lama Ahmad, Paul McMillan, Alex Beutel, Alexandre Passos & David G. Robinson, PRACTICES FOR GOVERNING AGENTIC AI SYSTEMS 1 (OpenAI Tech. Rep. 2023) (proposing governance practices for increasingly autonomous AI agent systems), https://openai.com/research/practices-for-governing-agentic-ai-systems [https://perma.cc/DG8H-K8Q9]; Rina Diane Caballar, *Explainer: What Are AI Agents?*, IEEE SPECTRUM (Nov. 19, 2024) (surveying the capabilities and applications of contemporary AI agents), https://spectrum.ieee.org/ai-agents [https://perma.cc/5TMV-6CFT].

[8] The distinction between prior scholarship and this Article turns critically on fluid agency. Foundational work, such as Samir Chopra & Laurence F. White, A LEGAL THEORY FOR AUTONOMOUS ARTIFICIAL AGENTS (2011), considers sophisticated but predictable tools that execute human-specified instructions with limited deviation (e.g., an e-commerce website). The causal chain from human command to tool action remains direct and traceable. *See id.* at 13-15. By contrast, when an AI interprets high-level goals, devises novel strategies, and executes multi-step plans, it may deviate substantially from—or even operate without—human instruction. This obscures the link between human intent and AI action. *See Fluid Agency*, *supra* note 7, Part I. *See also* STUART RUSSELL, HUMAN COMPATIBLE: ARTIFICIAL INTELLIGENCE AND THE PROBLEM OF CONTROL 139-41 (2019) (distinguishing between AI systems optimized for explicitly programmed objectives and those that learn and pursue objectives through autonomous exploration); NICK BOSTROM, SUPERINTELLIGENCE: PATHS, DANGERS, STRATEGIES 127-29 (2014) (analyzing how advanced AI systems might pursue goals through unexpected and unforeseeable means).

[9] *See* Thaler v. Vidal, 43 F.4th 1207, 1213 (Fed. Cir. 2022) (holding that "only a natural person can be an inventor" and that AI systems are not "individuals" under the Patent Act), cert. denied, 143 S. Ct. 1783 (2023); Chopra & White, *supra* note 8, at 153–56 (defining legal personhood as the capacity to be a subject of legal rights and obligations, drawing on John Chipman Gray's classic formulation). *See also* Idaho Code Ann. § 5-346 (2022) ("artificial intelligence . . . shall not be granted personhood"); Utah Code Ann. § 63G-32-102 (2024) (same).

liability.[10] Modern AI is neither. It is a partially autonomous system that increasingly operates in complex, multi-agent and human-agent environments.[11] Its operational independence *veils* human choices, even as its lack of personhood prevents it from bearing responsibility, creating novel *responsibility gaps* while amplifying existing ones.

These structural features create critical doctrinal complications. For instance, suppose an end-user tasks an AI ("CreatorBot") with a facially lawful and general goal, such as "generate a comprehensive report on market trends." Pursuing this objective, CreatorBot autonomously designs and deploys multiple subordinate AI agents ("ScraperBots") to simultaneously harvest vast quantities of proprietary, copyrighted information, thereby committing widespread infringement.[12]

---

[10] *See* RESTATEMENT (THIRD) OF TORTS: PRODS. LIAB. §§ 1–2, 19(a) (Am. L. Inst. 1998) (imposing liability on commercial sellers for harm caused by defective "tangible personal property distributed commercially for use or consumption"); RESTATEMENT (THIRD) OF TORTS: LIAB. FOR PHYSICAL & EMOTIONAL HARM § 7 (Am. L. Inst. 2010) (imposing a duty of reasonable care on any "actor" whose conduct creates a risk of physical harm). The former governs low-autonomy instrumentalities through the chain of production and sale; the latter governs persons whose own conduct causes injury. Neither framework contemplates a partially autonomous nonperson that independently generates risk. *See* Andreas Matthias, *The Responsibility Gap: Ascribing Responsibility for the Actions of Learning Automata*, 6 Ethics & Info. Tech. 175, 175–77 (2004) (arguing that autonomous learning machines create situations in which traditional responsibility ascription to the manufacturer or operator breaks down).

[11] Tool use is a key distinction between traditional generative AI and modern AI. For instance, while generative models can only produce a plan, modern AI can autonomously execute a plan. This is often achieved by combining a large language model with "action" modules, enabling the AI to interact with its digital and physical environment. For example, AIs can spawn other AIs, access resources managed by other AI, and coordinate with humans via communications APIs. By 2026, these capabilities had become routine in commercial AI developer tools, where a user's single high-level instruction can cause the system to autonomously spawn teams of specialized sub-agents for research, code analysis, file operations, and other tasks. *See generally* Lilian Weng, *LLM Powered Autonomous Agents*, LIL'LOG (June 23, 2023), https://lilianweng.github.io/posts/2023-06-23-agent/ [https://perma.cc/Y332-H3ED] (providing an overview of agent architecture); SHAVIT ET AL., *supra* note 7, at 3-5 (describing AI systems' ability to use tools to take long sequences of actions); Shunyu Yao, Jeffrey Zhao, Dian Yu, Nan Du, Izhak Shafran, Karthik Narasimhan & Yuan Cao, *ReAct: Synergizing Reasoning and Acting in Language Models*, *in* PROC. 11TH INT'L CONF. ON LEARNING REPRESENTATIONS (2023) (introducing the influential "Reason-Act" framework that explicitly combines verbal reasoning with action execution via tools).

[12] The capacity for AIs to autonomously spawn and manage multiple sub-agents has progressed rapidly from experiment to commercial reality. By 2026, leading providers had integrated autonomous sub-agent orchestration into standard commercial developer tools. *See, e.g.,*

When harm occurs, the law confronts a structural void. The directly infringing ScraperBots, lacking juridical capacity, cannot be sued.[13] CreatorBot, lacking juridical capacity, cannot be sued.[14] And the relevant human actors are insulated by AI's autonomy.[15] The user's general prompt fails to establish direct liability because the AI's autonomous decision-making process—and the multiplicity of subordinate autonomous agents it deploys—*obscures the causal*

---

Anthropic, *Introducing Claude Opus 4.6* (Feb. 5, 2026), https://www.anthropic.com/news/claude-opus-4-6 [https://perma.cc/U3R3-A3RK] (describing autonomous "agent teams" that work in parallel on complex tasks); GOOGLE, *Sub-agents*, GEMINI CLI DOCUMENTATION, https://geminicli.com/docs/core/subagents/ [https://perma.cc/L478-K6ER] (last visited Feb. 6, 2026) (describing sub-agents that execute tools without individual user confirmation); OPENAI, *Agents SDK*, OPENAI DEVELOPER DOCUMENTATION, https://developers.openai.com/codex/guides/agents-sdk/ [https://perma.cc/USS9-7Q6D] (last visited Feb. 6, 2026) (describing multi-agent orchestration with configurable approval policies, including fully autonomous execution).

[13] The inability to name a non-person as a defendant is a bedrock principle of tort law. An action for harm caused by an animal, for instance, is brought against its owner or keeper, not the animal itself, which is treated as property lacking juridical capacity. *See, e.g.,* RESTATEMENT (THIRD) OF TORTS: LIAB. FOR PHYSICAL & EMOTIONAL HARM §§ 21-23 (Am. L. Inst. 2010); Cetacean Cmty. v. Bush, 386 F.3d 1169, 1176 (9th Cir. 2004) (noting that "[i]t is obvious that an animal cannot function as a plaintiff in the same manner as a juridically competent human being," a principle that applies with equal force to its capacity to be a defendant). This principle of non-personhood has been consistently applied in analogous high-technology contexts, where courts have refused to grant legal status to AI systems. *See, e.g., Thaler v. Vidal*, 43 F.4th at 1213; Naruto v. Slater, 888 F.3d 418, 426 (9th Cir. 2018) (holding that animals lack statutory standing under the Copyright Act, reinforcing the principle that non-persons cannot invoke legal processes). Legal scholars have long recognized that under current law, an AI is considered property and thus cannot be a party to a lawsuit. *See, e.g.,* F. Patrick Hubbard, *"Sophisticated Robots": Balancing Liability, Regulation, and Innovation*, 66 FLA. L. REV. 1803, 1825-26 (2014) (explaining that "current law treats the robot as a machine" that cannot itself bear liability); Joanna J. Bryson, Mihailis E. Diamantis & Thomas D. Grant, *Of, For, and by the People: The Legal Lacuna of Synthetic Persons*, 25 ARTIF. INTELL. & L. 273 (2017).

[14] *See supra* note 13.

[15] *See Fluid Agency*, *supra* note 7, Parts I and II (describing how AI's stochastic, dynamic, and adaptive behavior creates a "fundamental unmappability" that fractures the causal link between human instruction and AI action); Matthias, *supra* note 10, at 177–79 (arguing that when autonomous systems learn and adapt beyond their initial programming, the manufacturer or operator "is in principle not capable of predicting the future machine behaviour," undermining traditional responsibility ascription). *See also* Chopra & White, *supra* note 8, at 107–18 (analyzing the difficulty of attributing an artificial agent's knowledge and actions to its principal under agency doctrine when the agent acts beyond the scope of its instructions).

*link* between the general instruction and the specific infringing acts; the law requires a volitional act of copying, but the AI's agency has rendered that human act too remote and general to ground liability.[16] Theories of secondary liability fare no better. Contributory liability falters where a general instruction provides no evidence of specific knowledge or willful blindness.[17] Vicarious liability becomes unavailable where the user lacks the "right and ability to supervise" the autonomous actions that *insulate the process from direct oversight*.[18] And a developer may invoke the Sony substantial-noninfringing-uses principle, arguing that an AI is a general-purpose tool with substantial noninfringing uses.[19]

For the relevant human actors—the user, deployer, developer—the AI's autonomy becomes a shield as it *obscures the foreseeability and intent* on which traditional doctrines depend.[20] While even a single fully autonomous machine can create a "responsibility gap," *multi-agent* AI environments magnify this issue into an *accountability chasm*—a fundamental inability to trace culpability through the system—creating a crisis of attribution that demands a new analytical approach.[21]

This Article introduces *'Operational Agency' (OA)* to bridge this chasm. OA neither anthropomorphizes AI nor advocates for legal personhood. Rather, it is proposed as a structured evidentiary tool—a *permeable legal fiction*—to pierce the veil of autonomy, allowing us to 'see through' an AI's operations to the human choices that shaped them. This is possible because, while lacking consciousness and culpability, modern AI is *emulative of human agency*; its behavior derives from processes optimized for reward signals aligned with human decision-making patterns.[22] By examining an AI's observable operational characteristics—*goal-directedness* (as a proxy for intent), *predictive processing* (as a proxy for foresight), and *safety architecture* (as a proxy for standard of care)—we can analyze its emulated agency. To operationalize that analysis, this Article also introduces *Operational Agency Graph (OAG)*, a methodological tool to visually map complex causal human-AI and AI-AI interactions, to systematically establish culpability.

---

[16] *See* Religious Tech. Ctr. v. Netcom On-Line Commc'n Servs., Inc., 907 F. Supp. 1361, 1370 (N.D. Cal. 1995). *Accord* Cartoon Network LP v. CSC Holdings, Inc. (Cablevision), 536 F.3d 121, 131-33 (2d Cir. 2008).

[17] *See* A&M Recs., Inc. v. Napster, Inc., 239 F.3d 1004, 1020-21 (9th Cir. 2001) [hereinafter *Napster*].

[18] *See* Perfect 10, Inc. v. Amazon.com, Inc., 508 F.3d 1146, 1173-75 (9th Cir. 2007).

[19] *See* Sony Corp. of Am. v. Universal City Studios, Inc., 464 U.S. 417, 442 (1984).

[20] The challenge that AI's unpredictability poses to the legal standard of foreseeability is a central theme in AI and law scholarship. *See, e.g.,* Ryan Calo, *Robotics and the Lessons of Cyberlaw*, 103 CAL. L. REV. 513, 550-51 (2015) (explaining that the emergent behavior of complex robotic systems challenges legal frameworks built on foreseeability); Yavar Bathaee, *The Artificial Intelligence Black Box and the Failure of Intent and Causation*, 31 HARV. J.L. & TECH. 889, 915-19 (2018) (arguing that the opacity of advanced AI systems makes it difficult to establish foreseeability and causation); Matthew U. Scherer, *Regulating Artificial Intelligence Systems: Risks, Challenges, Competencies, and Strategies*, 29 HARV. J.L. & TECH. 353, 365-68 (2016) (describing unpredictability as a key regulatory challenge for AI).

[21] *See* Matthias, *supra* note 10. In prior work, we establish *functional equivalence* as a guiding principle. *See Fluid Agency*, *supra* note 7, Part VI. Here, we provide a granular evidentiary framework to make it operational.

The discussion proceeds as follows. *Part I* demonstrates how core doctrines, from *mens rea* to products liability, break down when confronted with fluid agency. *Part II* responds by defining OA and its three analytical pillars. *Part III* introduces OAG, a methodological tool for mapping complex causal webs across human and AI actors. Part IV applies the framework across five case studies in tort, civil rights, constitutional law, and antitrust. Part V shows how OA integrates with established liability doctrines, functioning as both sword and shield. Part VI explores broader implications for law, policy, and governance. *Part VII* concludes.

## I. Fractured Foundations: Agency Without Personhood Breaks Core Doctrines

At the root of these doctrinal failures is the law's consistent refusal to grant legal status to non-natural actors—no matter how autonomous—that fall outside the established categories of natural and juridical persons. Federal Rules of Civil Procedure, for instance, treat the capacity to sue or be sued as a threshold question, and federal criminal law is grounded in the concept of a legal "person" who can be charged with a crime.[23] Courts have consistently refused to expand these categories, holding, for example, that animals lack standing to sue for environmental harms because they lack the requisite juridical capacity.[24] This long-standing principle applies with equal force to AI. While its most visible recent affirmations are in intellectual property, where AI systems cannot be inventors or authors, the principle is nearly universal under current law: legally, an AI is indistinguishable from any other piece of property, a mere instrument.[25]

[22] This is the cumulative effect of the doctrinal failures described *supra* notes 5–8 and accompanying text. Each theory fails for the same structural reason: the AI's autonomous decision-making interposes sufficient causal and epistemic distance between the human actor and the infringing conduct to defeat the requisite mental state. *See Fluid Agency*, *supra* note 7, Part II (describing how fluid agency's "fundamental unmappability" fractures doctrines that assume traceable provenance between human instruction and resulting conduct); Matthias, *supra* note 10, at 177–79. For an analogous shielding effect in the corporate context, see Coffee, *supra* note 5, at 389–93 (describing how corporate structure diffuses individual culpability).

[23] *See* Fed. R. Civ. P. 17(b). *See, e.g.,* 1 U.S.C. § 1 (defining "person" and "whoever" to include "corporations, companies, associations, firms, partnerships, societies, and joint stock companies, as well as individuals").

[24] *See Cetacean Cmty.*, 386 F.3d at 1176 (holding that animals lack standing because they are not "persons" under the relevant statutes).

Yet even as the law maintains the wall between human and machine, denying legal agency to what it deems tools, technology closes the distance. Fluid agency—the capacity of modern AI to interpret goals, devise strategies, and execute multi-step plans—emulates human agency but without consciousness or culpability.[26] It *veils the human intent and control* on which culpability depends: doctrines forged over centuries are forced to operate in a world where the immediate actor is, by definition, legally absent. This Part demonstrates how that structural mismatch fractures core tenets of criminal law (*mens rea*), civil liability (volitional conduct and products liability), agency law, and even administrative and constitutional due process—a breakdown made vivid by a wave of recent litigation.

The tension first becomes apparent in doctrines governing individual liability. A central pillar of criminal law is the requirement of a guilty mind, or *mens rea*. For liability to attach, a prosecutor must typically prove not only that the defendant committed a wrongful act (*actus reus*) but also that they did so with a culpable mental state, such as intent, knowledge, or recklessness.[27] However, while modern AI can autonomously perform an action that constitutes an *actus reus*—such as accessing a protected computer system or causing a vehicle collision—as a non-sentient computational entity, it cannot possess the requisite *mens rea*.[28] Fluid agency *inserts a non-culpable decision-making process that obscures intent*, while its lack of legal

---

[25] *See Thaler*, 43 F.4th at 1213; U.S. COPYRIGHT OFFICE, COMPENDIUM OF U.S. COPYRIGHT OFFICE PRACTICES § 306 (3d ed. 2021).

[26] *See Fluid Agency*, *supra* note 7, Parts I and II (defining fluid agency as behavior that is stochastic, dynamic, and adaptive, and arguing that it irreducibly entangles human and machine contributions).

[27] *See, e.g.,* MODEL PENAL CODE § 2.02 (Am. L. Inst. 1962) (defining general requirements of culpability). While the law has imposed criminal liability on corporations, an entity that cannot have a mental state, under the doctrine of *respondeat superior*, those frameworks still presuppose a culpable human agent whose intent can be imputed to the entity. *See* N.Y. Cent. & Hudson River R.R. Co. v. United States, 212 U.S. 481, 494-95 (1909).

[28] *See, e.g.,* 18 U.S.C. § 1030(a)(2), (a)(5) (2018) (criminalizing whoever "intentionally accesses a computer without authorization" or "knowingly causes the transmission of a program . . . and as a result of such conduct, intentionally causes damage"—elements whose mental-state requirements an AI system cannot satisfy); *see also* Cal. Penal Code § 192(c) (West 2024) (defining vehicular manslaughter as driving "in the commission of an unlawful act . . . or . . . with gross negligence"— a standard that presupposes a human driver capable of negligence).

agency means that there is no culpable "mind" to which criminal intent can attach at the point of action.[29]

Consequently, in systemic failures, liability may be deflected onto the most proximate and legally cognizable human actor, regardless of their actual control over the AI's actions.[30] In a 2018 fatal incident involving an Uber self-driving test vehicle, prosecutors declined to bring criminal charges against the corporation for the actions of its autonomous system, concluding there was "no basis for criminal liability for the Uber corporation" itself.[31] Instead, prosecutors charged the human safety driver, to whom *mens rea* and *actus reus* could attach.[32]

Neither is the *actus reus* always easy to isolate. Modern AI is inherently opaque; its internal decision-making pathways can be inscrutable, even to its own developers, making it extraordinarily difficult to retrospectively determine why an AI behaved in a particular way.[33] In large-scale multi-agent environments, an "act" can be fragmented across thousands of algorithmic micro-decisions such that the legally relevant conduct becomes visible only in the aggregate.[34] For instance, an algorithmic cartel may emerge only when the combined outputs of many pricing bots are viewed together.[35] In such scenarios, the law may find neither a mind to probe for intent nor communication to trace back to a legal actor's decision (as is typical in antitrust litigation with human decision-makers). The opacity and complexity can effectively shield the *actus reus*.[36]

---

[29] This principle is reflected in judicial reasoning, scholarly consensus, and a consistent pattern of prosecutorial practice. *See, e.g.,* United States v. Athlone Indus., Inc., 746 F.2d 977, 979 (3d Cir. 1984) ("Since robots cannot be sued, but they can cause devastating damage, the defendant . . . was twice sued as the ultimate responsible distributor . . . ."); Gabriel Hallevy, WHEN ROBOTS KILL: ARTIFICIAL INTELLIGENCE UNDER CRIMINAL LAW ch. 2 (2013) (synthesizing doctrine to conclude that robots lack a "culpable mental state"); State v. Vasquez, No. CR2020-001853 (Ariz. Super. Ct. July 28, 2023) (involving the safety driver for an Uber autonomous test vehicle who pleaded guilty to endangerment); People v. Riad, No. TA155613 (L.A. Cnty. Super. Ct. June 22, 2023) (involving a Tesla driver who, with Autopilot engaged, entered a no-contest plea to vehicular manslaughter).

[30] *See* Madeleine Clare Elish, *Moral Crumple Zones: Cautionary Tales in Human–Robot Interaction*, 5 ENGAGING SCI. TECH. & SOC'Y 40 (2019).

[31] While the prosecutor's office did not explicitly invoke the concept of *mens rea*, its conclusion effectively treated the autonomous system as incapable of forming criminal intent. *See* Letter from Sheila Sullivan Polk, Yavapai Cnty. Att'y, to Maricopa Cnty. Att'y's Office (Mar. 4, 2019) (explaining the decision not to bring criminal charges against Uber).

[32] *See* NAT'L TRANSP. SAFETY BD., COLLISION BETWEEN VEHICLE CONTROLLED BY DEVELOPMENTAL AUTOMATED DRIVING SYSTEM AND PEDESTRIAN, NTSB/HAR-19/03, at v, 60 (2019) (finding the probable cause of the crash to be the human operator's failure to monitor the environment due to distraction).

[33] *See* Bathaee, *supra* note 20, at 903-15.

[34] *See Fluid Agency*, *supra* note 7, Parts I and II (describing how fluid agency distributes causally relevant conduct across stochastic, dynamic, and adaptive processes such that human and machine contributions become "irreducibly entangled"); Chan et al., *supra* note 6, at 960

A similar strain is evident in civil liability. In copyright law's "volitional conduct" doctrine, the failure shifts from an "actor's mind" to the "actor's hand." To hold a defendant directly liable, courts require proof of a volitional act—an affirmative step that pins responsibility on the person who, in effect, "pushes the copy button."[37] However, AI severs the causal chain. For instance, in our prior hypothetical, the user's general prompt is separated from the infringing act by the AI's autonomous process; it is CreatorBot and the many ScraperBot sub-agents, not the user, that select the source and execute the unauthorized copy. The volitional conduct doctrine falters because, in searching for a human actor and a human-initiated act of fixation, it finds only algorithms autonomously executing a self-directed operational process.

---

(analyzing how interactions between many deployed agents can produce systemic harms not present at the level of any single system, including "destabilising feedback loops" and "unpredictable behavioural changes that are characteristic of complex systems").

[35] *See* Emilio Calvano, Giacomo Calzolari, Vincenzo Denicolò & Sergio Pastorello, *Artificial Intelligence, Algorithmic Pricing, and Collusion*, 110 AM. ECON. REV. 3267 (2020) (finding experimentally that AI-powered pricing algorithms consistently learn to charge supracompetitive prices without explicit communication or a pre-programmed collusive agreement).

[36] This opacity parallels due process concerns in algorithmic adjudicative settings. *See* State v. Loomis, 881 N.W.2d 749, 769-71 (Wis. 2016) (holding that a defendant's inability to challenge the proprietary COMPAS algorithm did not violate due process, but setting forth specific cautionary conditions for its use in sentencing to ensure transparency and accuracy); *see also* Note, *State v. Loomis: Wisconsin Supreme Court Requires Warning Before Use of Algorithmic Risk Assessments in Sentencing*, 130 HARV. L. REV. 1530, 1531 (2017) (criticizing the court's remedy as an "ineffective means" of protecting due process rights and noting that the required warnings fail to address the inability of judges to properly evaluate the secret methodologies of risk assessment tools).

[37] The principle that direct liability requires a volitional act was famously articulated in *Religious Technology Center v. Netcom On-Line Communication Services, Inc.*, 907 F. Supp. at 1370, and is now a cornerstone of the doctrine. *See, e.g., Cartoon Network LP*, 536 F.3d at 131; CoStar Grp., Inc. v. LoopNet, Inc., 373 F.3d 544, 550 (4th Cir. 2004); Capitol Records, LLC v. ReDigi Inc., 910 F.3d 649, 660 (2d Cir. 2018) (holding that ReDigi's transfer process created new reproductions, constituting direct infringement and foreclosing a first-sale defense); Perfect 10, Inc. v. Giganews, Inc., 847 F.3d 657, 666 (9th Cir. 2017) (rejecting direct liability where a system "responded automatically to user commands" because the human volitional element was missing).

The logical alternative, strict products liability, proves equally unavailing. Reasoning that intangible software (such as AI) provides only "information, guidance, ideas, and recommendations," courts often refuse to classify it as a "product" under the Restatement,[38] instead channeling plaintiffs toward negligence claims.[39] Nevertheless, these negligence claims can flounder on the same fundamental obstacle: an AI's agency can obscure the foreseeability needed to prove a breach of care.

For instance, in October 2023, a Cruise autonomous vehicle's post-collision "pullover" maneuver dragged a trapped pedestrian, creating a liability puzzle.[40] With no human driver to blame, the inquiry had to locate liability not in a single volitional act, but in the product's

---

[38] *See* Rodgers v. Christie, 795 F. App'x 878, 880 (3d Cir. 2020); Winter v. G.P. Putnam's Sons, 938 F.2d 1033 (9th Cir. 1991) (holding that an informational book is not a "product" for strict liability purposes); RESTATEMENT (THIRD) OF TORTS: PRODS. LIAB. § 19 cmt. d (Am. L. Inst. 1998).

[39] This traditional firewall may have begun to fracture. In *Garcia v. Character Technologies, Inc.*, the court refused to dismiss a products-liability claim against an AI chatbot provider, drawing a critical distinction between the chatbot's expressive "content" and its functional "design". *See* Order at 35, Garcia v. Character Techs., Inc., No. 6:24-cv-01903-ACC-UAM (M.D. Fla. May 21, 2025), ECF No. 115, https://storage.courtlistener.com/recap/gov.uscourts.flmd.433581/gov.uscourts.flmd.433581.115.0.pdf [https://perma.cc/2BDF-T6YH]. However, this exception may have limited application to purely functional AI agents, which lack the anthropomorphic design features found to be at stake in *Garcia*.

[40] Investigation into the Cruise incident revealed the failure was not one of perception but of logic. The autonomous vehicle's sensors detected and tracked the pedestrian, but its software misclassified the event, failing to recognize the pedestrian was pinned beneath it. This led the AI to execute its default "pullover" maneuver, a pre-programmed safety routine that proved catastrophic because its logic did not adequately account for this specific, high-risk post-collision scenario. These key facts are documented across multiple official sources. *See* CAL. DEP'T OF MOTOR VEHICLES, ORDER OF SUSPENSION TO CRUISE LLC (Oct. 24, 2023) , https://www.dmv.ca.gov/portal/news-and-media/dmv-statement-on-cruise-llc-suspension/ [https://perma.cc/TF24-UAG8]; QUINN EMANUEL URQUHART & SULLIVAN, LLP, REPORT TO THE BOARDS OF DIRECTORS REGARDING THE OCTOBER 2, 2023 ACCIDENT 1, 12-15 (Jan. 24, 2024), https://assets.ctfassets.net/95kuvdv8zn1v/1mb55pLYkkXVn0nXxEXz7w/9fb0e4938a89dc5cc09bf39e86ce5b9c/2024.01.24_Quinn_Emanuel_Report_re_Cruise.pdf [https://perma.cc/HC5T-A8GF] [hereinafter QUINN EMANUEL REPORT]; *see also* NAT'L HIGHWAY TRAFFIC SAFETY ADMIN., PART 573 SAFETY RECALL REPORT NO. 23E-086 2 (Nov. 7, 2023) https://static.nhtsa.gov/odi/rcl/2023/RCLRPT-23E086-7725.PDF [https://perma.cc/E8SP-7XCF] [hereinafter NHTSA SAFETY RECALL REPORT] (describing how the software "inaccurately characterized the collision as a lateral collision" and commanded an improper "pull over" maneuver). This example is explored in detail in Part IV.

architecture—a task for which traditional negligence analysis is ill-equipped: any user of an autonomous vehicle can point to their general, facially lawful instructions, and any developer to their general-purpose tool. For both actors, AI autonomy obscures not only foreseeability but standard of care itself, insulating them from the fault-based inquiry that negligence demands.

Similar strains appear in doctrines governing organizational and systemic liability. Traditional agency law, which attributes the actions of an agent to a principal, offers a potential path to accountability, yet it too faces conceptual hurdles. While AI systems are frequently described as "agents," they cannot form a legal agency relationship, which requires consent, fiduciary duties, and a shared understanding between two legal persons.[41] More fundamentally, the very fluid agency that makes AI powerful *obscures the lines of control and foreseeability* that traditionally underpin vicarious liability. The AI is not merely a passive conduit for human commands; it interprets, strategizes, and may generate novel pathways to achieve its objectives, thereby *insulating its human "principal" from direct supervision*.[42] The analogy is tempting but breaks down because the AI's fluid agency weakens the principal's direct control, while its lack of legal agency means it cannot be a true agent. The doctrine collapses because the essential elements of control and legal status are simultaneously absent.

This collapse creates a novel *systemic enterprise responsibility gap*, where an organization can deploy and benefit from an autonomous system whose very nature shields the corporation from traditional forms of vicarious liability.[43]

The breakdown of private law is matched by a parallel fracture in public and administrative law, as fluid agency erodes the fundamental right to procedural due process. The Constitution guarantees individuals notice and a meaningful opportunity to be heard before the government deprives them of life, liberty, or property.[44] But when governments or their agents rely on opaque algorithms for critical decisions, citizens may be left unable to understand—let alone challenge—the basis for state action against them. The "opportunity to be heard" becomes an empty formality when the affected individual cannot interrogate the reasoning that determines their fate.

---

[41] *See* RESTATEMENT (THIRD) OF AGENCY § 1.01 (Am. L. Inst. 2006) (defining agency as "the fiduciary relationship that arises when one person (a 'principal') manifests assent to another person (an 'agent') that the agent shall act on the principal's behalf and subject to the principal's control, and the agent manifests assent or otherwise consents so to act").

[42] *See* David C. Vladeck, *Machines Without Principals: Liability Rules and Artificial Intelligence*, 89 WASH. L. REV. 117, 141-45 (2014) (discussing the limitations of applying traditional agency principles to autonomous systems).

[43] *See Fluid Agency*, *supra* note 7, Part V (analyzing how fluid agency destabilizes traditional liability frameworks and identifying a "systemic enterprise responsibility gap" where organizations may evade liability for harms caused by autonomous systems they deploy).

[44] *See* Mathews v. Eldridge, 424 U.S. 319, 333 (1976).

Courts have grappled with such due process failures across multiple contexts. In *Houston Federation of Teachers*, a federal court found that the Houston school district violated teachers' procedural due process rights by terminating them based on a proprietary algorithm. Because the system's methodology was secret and its outputs could not be verified or challenged, the teachers were "denied any meaningful opportunity to ensure correct calculation of their scores"—the very deprivation of a fair hearing that procedural due process forbids.[45] In *State v. Loomis*, the Wisconsin Supreme Court confronted a similar challenge to the use of COMPAS, a proprietary risk-assessment algorithm used in criminal sentencing. The court declined to find a due process violation but recognized the constitutional concern, imposing specific cautionary conditions: sentencing courts must be warned that the algorithm's proprietary nature prevents individualized assessment, that its accuracy has not been validated for the defendant's specific circumstances, and that it must not be used as the determinative factor.[46] Commentators, however, criticized this remedy as inadequate—a set of warnings that "ignores judges' inability to evaluate risk assessment tools."[47]

Moreover, even when defendants attempt to sidestep these doctrinal failures, they encounter new manifestations of the same fundamental paradox. Increasingly, AI developers are invoking the First Amendment as a shield, arguing that their systems' outputs constitute protected speech that cannot be the basis for tort liability. If accepted, this defense would subject content-based liability claims to strict scrutiny—a formidable legal barrier.[48] Yet, in *Garcia v. Character Technologies, Inc.*, the court refused to dismiss claims on First Amendment grounds, questioning whether an LLM's output can be "expressive" in the constitutional sense when no human has made an "inherently expressive choice."[49] This judicial skepticism reveals how the accountability chasm extends even to constitutional defenses: the AI's output is too autonomous to be attributed to human expression, yet too synthetic to claim the protections reserved for speakers with constitutional rights.

---

[45] *See* Houston Fed'n of Teachers, Local 2415 v. Houston Indep. Sch. Dist., 251 F. Supp. 3d 1168, 1176, 1179 (S.D. Tex. 2017).

[46] *See Loomis*, 881 N.W.2d at 769-71.

[47] *See* Note, *supra* note 36, at 1531.

[48] Under First Amendment doctrine, content-based restrictions on speech are presumptively unconstitutional and subject to strict scrutiny. *See* Reed v. Town of Gilbert, 576 U.S. 155, 163 (2015). This would subject content-based tort claims against AI providers to a formidable, though not insurmountable, legal barrier.

[49] *See Garcia*, No. 6:24-cv-01903-ACC-UAM, at 30-31. The court distinguished between algorithms that "simply implement [the entity's] inherently expressive choice" and AI systems where humans "hand the reins to an [AI] tool." *Id.* at 31 (quoting Justice Barrett's concurrence in Moody v. NetChoice, LLC, 603 U.S. 707, 746 (2024) (Barrett, J., concurring)). The court concluded it was "not prepared to hold that Character A.I.'s output is speech" at the motion-to-dismiss stage. *Id.*

These collective limitations—the inapplicability of *mens rea*, the erosion of clear volitional acts, the foreclosure of strict liability, the breakdown of due process, the strained analogies in agency law, the refusal to grant legal status to non-humans, the evidentiary fog of "black box" systems, and the uncertain applicability of constitutional protections—converge to create the accountability chasm. Across criminal, civil, administrative, and constitutional domains, the foundational failure remains the same: the AI's fluid agency inserts a non-liable, decision-making layer that *obscures the chain of causation and intent*, shielding the human actors who designed, deployed, and benefited from the system. Having established the fractured foundations of current doctrine, our analysis now turns from diagnosing the problem to proposing a solution: the Operational Agency framework.

## II. Bridging the Chasm: The Operational Agency Framework

The accountability chasm forces a choice between two competing legal fictions. One path is to create an *opaque* fiction by granting AI systems legal agency—in effect, a form of legal personhood—allowing the machine to bear rights and duties. Yet, as the following discussion reveals, this approach remains the subject of intense debate as it risks the creation of new liability shields for human actors. The alternative, which this Article advances, is a new *permeable* legal fiction: a structured evidentiary framework to trace liability back to the persons whose choices the system operationalizes.

Proponents of AI personhood argue that conferring legal agency on AI could fill accountability gaps and encourage responsible innovation. Some, including former federal judges, have recently distilled the doctrinal and ethical arguments for extending at least some legal personhood to advanced AI.[50] Others have proposed treating advanced AI like a corporate entity with limited liability, where the AI itself would be a legal "person" required to carry insurance to compensate victims, while its human creators face liability only for egregious conduct.[51] Still others note that legal personality is "divisible," suggesting specific rights and duties could be tailored to AI without granting full human rights.[52] This idea gained political traction when the European Parliament recommended exploring a special legal status for the "most sophisticated autonomous robots," so they could be deemed "electronic persons" responsible for any damage they may cause,[53] and some scholars have demonstrated that current business-entity law can be structured so an algorithm exercises *de facto* control over a limited liability company.[54]

[50] *See, e.g.,* Katherine B. Forrest, *The Ethics and Challenges of Legal Personhood for AI*, 133 YALE L.J. FORUM 1175 (2024).

[51] *See, e.g.,* Alicia Lai, *Artificial Intelligence, LLC: Corporate Personhood as Tort Reform*, 2021 MICH. ST. L. REV. 597.

[52] *See* Paulius Čerka, Jurgita Grigienė & Gintarė Sirbikytė, *Is it possible to grant legal personality to artificial intelligence software systems?*, 33 COMPUT. L. & SEC. REV. 685 (2017).

Conversely, critics oppose AI personhood, warning of severe legal and moral pitfalls. They argue that creating "electronic persons" is "morally unnecessary and legally troublesome," as it would likely create a "legal black hole" that absorbs human responsibilities but cannot be truly punished or held to account.[55] Official advisory bodies within the European Union share this view, explicitly rejecting the Parliament's proposal and warning of moral-hazard and liability-evasion risks.[56] The primary risk is that an AI "person" would become a convenient liability shield, allowing human developers and deployers to deflect blame for the harms their systems cause.[57] Likewise, comprehensive reviews of legal personhood doctrine conclude that current AI systems do not qualify as legal persons under any principled criteria and that extending such status is incompatible with established principles of legal accountability.[58] A central worry is that an AI cannot fulfill the obligations of personhood; it cannot be jailed, and it may be judgment-proof if it lacks assets beyond what its creators provided.[59] Ultimately, these critics argue, AI should remain a tool under human control, not an independent legal actor that could be exploited for anti-social purposes.[60]

---

[53] While influential, this proposal was not adopted by the European Commission, which, along with other EU advisory bodies, later rejected the idea of AI personhood. *See infra* note 56.

[54] *See* Shawn Bayern, *The Implications of Modern Business-Entity Law for the Regulation of Autonomous Systems*, 19 STAN. TECH. L. REV. 93, 98 (2015).

[55] *See* Bryson, Diamantis & Grant, *supra* note 13, at 274, 289.

[56] *See, e.g.,* Opinion of the European Economic and Social Committee on Artificial Intelligence, 2017 O.J. (C 288) 1, ¶ 1.12 (concluding that creating a specific legal status for robots is "not appropriate" and warning against the moral hazard of such an approach); HIGH-LEVEL EXPERT GRP. ON AI, EUR. COMM'N, POLICY AND INVESTMENT RECOMMENDATIONS FOR TRUSTWORTHY AI 41 (2019) (recommending that liability for AI systems be attributed to existing natural or legal persons and cautioning against the creation of a separate legal personality for AI).

[57] *See* Bryson, Diamantis & Grant, *supra* note 13, at 281; *see also* RYAN ABBOTT, THE REASONABLE ROBOT: ARTIFICIAL INTELLIGENCE AND THE LAW 155-58 (2020) (arguing against AI personhood and in favor of holding humans responsible for AI actions).

[58] *See* S. M. Solaiman, *Legal Personality of Robots, Corporations, Idols and Chimpanzees: A Quest for Legitimacy*, 25 ARTIF. INTELL. & L. 155 (2017); Nadia Banteka, *Artificially Intelligent Persons*, 58 HOUS. L. REV. 537 (2021).

[59] *See* Bryson, Diamantis & Grant, *supra* note 13, at 282.

[60] *See* Lynn M. LoPucki, *Algorithmic Entities*, 95 WASH. U. L. REV. 887, 888 (2018).

To resolve this impasse, this Article introduces *Operational Agency (OA)*, a structured evidentiary framework—a *permeable legal fiction*—[61]to reinforce human accountability *through* a systematic analysis of an AI's observable characteristics and behavior. Unlike corporate personhood, which creates an opaque barrier that can shield natural persons from direct liability, OA functions as a permeable fiction—one designed to "look through" the AI to the human choices that shaped it. The objective is not to personify AI, but to trace an evidentiary trail back to the relevant human actors. By treating an AI's functional characteristics as a surrogate for the human will that shaped them, OA provides a principled method to apportion culpability—whether intent, knowledge, recklessness, or negligence—without upending the human-centric foundations of our legal system.

---

[61] The concept of legal fiction as an analytical tool has deep roots in Anglo-American jurisprudence. *See* LON L. FULLER, LEGAL FICTIONS 9-10 (1967) (defining legal fictions as "either (1) a statement propounded with a complete or partial consciousness of its falsity, or (2) a false statement recognized as having utility"); Aviam Soifer, *Reviewing Legal Fictions*, 20 GA. L. REV. 871, 877 (1986) (describing legal fictions as "constructs that help judges reach decisions by providing a form through which to understand and organize reality"). The distinction between "opaque" and "permeable" fictions draws on the functional differences in how legal constructs operate. Corporate personhood exemplifies an opaque fiction that creates a legal entity with its own rights and duties, effectively blocking direct attribution to natural persons in many contexts. *See* Sanford A. Schane, *The Corporation is a Person: The Language of a Legal Fiction*, 61 TUL. L. REV. 563, 565 (1987) (analyzing how corporate personhood functions as a complete legal substitute for natural persons); *cf.* Elizabeth Pollman, *Reconceiving Corporate Personhood*, 2011 UTAH L. REV. 1629, 1641-43 (distinguishing between different theories of corporate personhood and their implications for liability). In contrast, veil-piercing doctrine demonstrates how courts employ analytical frameworks to look through legal forms when necessary to prevent fraud or injustice. *See* Robert B. Thompson, *Piercing the Corporate Veil: An Empirical Study*, 76 CORNELL L. REV. 1036, 1038 (1991) (finding courts pierce the veil to reach individual shareholders in close corporations approximately 40% of the time); Stephen M. Bainbridge, *Abolishing Veil Piercing*, 26 J. CORP. L. 479, 513-14 (2001) (describing veil-piercing as a "safety valve" that allows courts to balance entity protection with prevention of abuse). Our "permeable fiction" concept operates similarly by providing a structured method to "look through" an AI's fluid agency to the human choices that shaped it, though our concept functions not as an exceptional, equitable remedy applied to prevent fraud or injustice but as a primary, *ex post* evidentiary framework for assessing fault. This approach aligns with contemporary scholarship on legal fictions in technological contexts. *See* Meg Leta Jones, *Does Technology Drive Law? The Dilemma of Technological Exceptionalism in Cyberlaw*, 2018 U. ILL. J.L. TECH. & POL'Y 249, 276-78 (arguing that legal fictions help law adapt to technological change without abandoning core principles); CHOPRA & WHITE, *supra* note 8, at 155-83 (analyzing how legal fictions might apply to AI systems while ultimately cautioning against legal personhood and emphasizing the need to maintain human accountability).

To be legally coherent, any such framework must provide evidentiary proxies for the foundational components of legal fault. The accountability chasm identified in Part I arises because fluid agency fractures three irreducible linchpins of legal responsibility: a culpable mental state (*mens rea*), a foreseeable link to harm, and a breach of a standard of care. We therefore structure Operational Agency upon three analytical pillars, each calibrated to repair one of these fractures: (1) *Intent*, inferred from the AI's *Goal-Directedness*; (2) *Foreseeability*, established by its *Predictive Processing* capabilities; and (3) *Breach of Care*, assessed through its *Safety Architecture and Design Choices*.[62] We examine each of these pillars in depth below.

1. *Intent Through Goal-Directedness*

The first pillar provides a proxy for a culpable mental state (*mens rea*) by examining an AI's *goal-directedness*. The analysis is grounded in the established legal practice of inferring intent from objective evidence of purpose,[63] most notably the "collective knowledge" or "corporate scienter" doctrine used to establish corporate intent.[64]

---

[62] The tripartite structure of the OA framework reflects three irreducible elements of legal fault across both criminal and civil law: a wrongful mental state or negligent failure, a causal connection to harm, and a breach of a recognized duty. *See* H.L.A. HART & TONY HONORÉ, CAUSATION IN THE LAW 1-5 (2d ed. 1985) (identifying the essential elements of legal responsibility); MICHAEL S. MOORE, CAUSATION AND RESPONSIBILITY: AN ESSAY IN LAW, MORALS, AND METAPHYSICS 3-5 (2009) (analyzing the relationship between moral culpability and legal liability). The OA framework maps directly onto this established structure: goal-directedness provides evidence of mental state, predictive processing establishes the causal-foreseeability nexus, and safety architecture demonstrates adherence to or breach of the duty of care.

[63] The inference of intent from systematic, goal-oriented behavior has deep roots in both criminal and tort law. Courts routinely infer intent from patterns of conduct under the "natural and probable consequences" doctrine. *See* 1 WAYNE R. LAFAVE, SUBSTANTIVE CRIMINAL LAW § 5.2(a) (3d ed. 2017) (explaining that intent to cause a consequence exists when one "knows that those consequences are substantially certain to result from his acts"). This principle extends to organizational contexts, where systematic policies establish corporate intent even absent a "smoking gun" directive. *See also* Jennifer Arlen & Reinier Kraakman, *Controlling Corporate Misconduct: An Analysis of Corporate Liability Regimes*, 72 N.Y.U. L. REV. 687, 702-03 (1997) (describing "composite liability," where corporate intent is established from the collective conduct and knowledge of employees, even if no single employee is culpable). In the AI context, an algorithm's objective function represents an even more direct manifestation of purpose, as it mathematically defines what the system is optimized to achieve. *See* RUSSELL, *supra* note 8, at 173-74 (explaining that for a standard AI, its sole purpose is to optimize the fixed objective provided to it).

[64] Under this doctrine, the knowledge and purpose of multiple employees can be aggregated to form a single corporate "mind," even if no single individual possessed the full picture of wrongdoing. *See, e.g.,* United States v. Bank of New Eng., N.A., 821 F.2d 844, 856 (1st Cir.

An AI's goal-directedness is the algorithmic analogue of corporate policy. It is the encoded, operational will of its human creators that resides not in abstract motive but in discoverable technical artifacts, such as the design of a *reward function* in a reinforcement learning system, the hard-coded *objective function* in an optimization algorithm, or the implicit goals embedded in a *fine-tuning dataset*. Courts can thus distinguish between different levels of culpability by examining the objective purpose for which the system was designed and deployed.

For example, an AI whose core reward function is explicitly designed to circumvent copyright protections or to generate content that mimics a competitor's trade dress would provide direct evidence of *specific intent* to infringe. This satisfies the high bar for inducement liability set forth in *MGM Studios, Inc. v. Grokster, Ltd.*, which requires evidence of "an object of promoting its use to infringe copyright."[65] In contrast, a broader goal, such as "maximize user engagement at all costs," which foreseeably leads to the generation of harmful or addictive content, might point toward recklessness or negligence rather than specific intent. By scrutinizing the AI's engineered purpose, OA provides a principled, evidence-based method to map a system's operational priorities back to its developer's and deployer's culpable mental state.

2. *Foreseeability Through Predictive Processing*

The second pillar addresses knowledge and foreseeability by analyzing the AI's *predictive processing* capabilities—that is, its ability to model its environment and quantify likely outcomes before acting.[66] This pillar allows the law to establish *constructive knowledge* on the part of a human actor, preventing a developer from credibly claiming an outcome was unforeseeable when the very system they built and profited from was capable of predicting that outcome. Just as a pharmaceutical

1987) ("A collective knowledge instruction is entirely appropriate in the context of corporate criminal liability…. Corporations compartmentalize knowledge… [and] the aggregate of those components constitutes the corporation's knowledge….").

[65] *See* MGM Studios, Inc. v. Grokster, Ltd., 545 U.S. 913, 936-37 (2005) [hereinafter *Grokster*].

[66] The use of an AI system's own predictive capabilities to establish foreseeability finds strong precedent in products liability law, where manufacturers' internal risk assessments routinely serve as evidence of what hazards were known or knowable. *See, e.g., In re* Methyl Tertiary Butyl Ether ("MTBE") Prods. Liab. Litig., 175 F. Supp. 2d 593, 622 (S.D.N.Y. 2001) (finding that oil companies' internal documents acknowledging contamination risks established foreseeability); Toole v. Baxter Healthcare Corp., 235 F.3d 1307, 1314 (11th Cir. 2000) (holding that a manufacturer's internal study showing breast implant risks was admissible on the issue of notice); Grimshaw v. Ford Motor Co., 119 Cal. App. 3d 757, 775-79 (1981) (relying on internal cost-benefit analyses—the "Pinto Memos"—as evidence of the company's knowledge of fuel-system risks).

company cannot claim a side effect was unforeseeable when its own clinical data predicted it, an AI developer cannot claim ignorance of risks that their system's logs explicitly identified. This principle aligns with the emerging concept of "algorithmic foreseeability" in the AI governance literature.[67]

The inquiry focuses on two forms of machine-generated foresight. The first is the *"embedded knowledge"* contained within an AI's vast training data (e.g., a large language model trained has often ingested countless documents on copyright law and privacy norms).[68] While the precise contours of this imputation remain to be developed by courts, the principle is straightforward: a developer who trains a system on data reflecting known risks cannot later claim ignorance of those risks.[69] The second is the AI's functional capacity to generate *risk assessments*.[70] These assessments—in the form of confidence scores, logged warnings about potential policy violations, or similar discoverable artifacts—can constitute probative evidence (subject to authentication and applicable evidentiary rules) of what a reasonable actor possessing the same tool would have foreseen.[71]

---

[67] *See* Andrew D. Selbst, *Negligence and AI's Human Users*, 100 B.U. L. REV. 1315, 1342-45 (2020) (proposing that foreseeability analysis should incorporate what an algorithm itself can predict); Karni Chagal-Feferkorn, *The Reasonable Algorithm*, 2018 U. ILL. J.L. TECH. & POL'Y 111, 127-30 (arguing that an algorithm's capacity to process information should be imputed to its user for foreseeability analysis).

[68] "Embedded knowledge" refers to the patterns, facts, and relationships learned by a large language model from its training corpus, which often consists of a significant portion of the public internet. This data inevitably includes vast quantities of legally and socially salient information. *See, e.g.,* Tom B. Brown et al., *Language Models are Few-Shot Learners*, 33 ADV. NEURAL INFO. PROCESSING SYS. 1877, 1880 (2020) (describing the training data for GPT-3, which included the "Common Crawl" dataset and its billions of web pages); *see also* Mark A. Lemley & Bryan Casey, *Fair Learning*, 99 TEX. L. REV. 743, 748-51 (2021) (discussing how AI models ingest and learn from copyrighted and other legally protected materials as a core part of their function).

[69] *See* RESTATEMENT (THIRD) OF TORTS: PRODS. LIAB. § 2 cmt. m (Am. L. Inst. 1998) ("A seller is charged with knowledge of what reasonable testing would reveal."); Borel v. Fibreboard Paper Prods. Corp., 493 F.2d 1076, 1089 (5th Cir. 1973) ("The manufacturer is held to the knowledge and skill of an expert."). Training an AI system on a corpus is functionally a form of research; the developer should be charged with knowledge of risks that the training data reflects. *See also* Selbst, *supra* note 67, at 1342–45.

[70] Many advanced AI systems, particularly those offered as commercial services, include safety mechanisms that generate internal risk scores or flag outputs that may violate usage policies. These are discoverable technical artifacts. For example, OpenAI's Moderation API classifies content into categories and provides a confidence score for each. *See* OpenAI, *Moderation*, OPENAI API DOCS, https://platform.openai.com/docs/guides/moderation [https://perma.cc/3VVB-LTZG] (last visited July 15, 2025). Similarly, other safety-focused architectures are designed to evaluate outputs against a set of rules or principles before they are shown to the user. *See* Yuntao Bai et al., *Constitutional AI: Harmlessness from AI Feedback*, arXiv:2212.08073 (Dec. 15, 2022), https://arxiv.org/abs/2212.08073 [https://perma.cc/N4CW-

This analysis is crucial for establishing a *"reasonable AI developer"* standard of care, which holds professionals to the standard of a prudent expert in their field.[72] A prudent developer, guided by emerging industry standards, is expected to understand the predictive capacities and failure modes of their models, and required to act upon the knowledge their own systems generate.[73] If a system's own logs show it "predicting" that scraping copyrighted sites is the most efficient path to its goal, or that a certain conversational strategy carries a high risk of causing emotional distress, the developer's failure to prevent this action constitutes a failure to anticipate a foreseeable

---

WMMD] (describing a method for training an AI to supervise itself for safety, a process that inherently involves risk assessment).

[71] The principle that a company's internal documents, including safety reports and risk analyses, are admissible to prove knowledge or foreseeability is well-established in products liability law. The AI-generated logs and risk assessments described here are the modern digital equivalent of such documents. *See supra* note 66 (citing cases where manufacturers' internal risk assessments served as evidence of what hazards were known or knowable).'

[72] An actor who possesses superior skills or knowledge—such as a professional AI developer—is expected to exercise care commensurate with those capabilities. *See* RESTATEMENT (THIRD) OF TORTS: LIAB. FOR PHYSICAL & EMOTIONAL HARM § 12 (Am. L. Inst. 2010) (providing that an actor's skills or knowledge exceeding those possessed by most others are taken into account in determining whether the actor has behaved as a reasonably careful person).

[73] *See, e.g.,* NAT'L INST. OF STDS. & TECH., ARTIFICIAL INTELLIGENCE RISK MANAGEMENT FRAMEWORK (AI RMF 1.0) 19-20 (NIST AI 100-1, 2023) (emphasizing the need to map, measure, and manage risks throughout the AI lifecycle).

harm.[74] In this way, the system's own predictive architecture becomes the measure of its creator's constructive knowledge.[75]

3. *Breach of Care Through Safety Architecture and Design Choices*

The final pillar assesses the breach of a standard of care by scrutinizing the AI's *Safety Architecture and Design Choices*. This moves beyond explicit rules to evaluate the holistic reasonableness of the system's design, directly paralleling the "design defect" analysis in modern products liability law.[76] The conceptual inquiry is captured by

---

[74] *See The T.J. Hooper*, 60 F.2d 737, 740 (2d Cir. 1932) (Hand, J.) ("[I]n most cases reasonable prudence is in fact common prudence; but strictly it is never its measure; a whole calling may have unduly lagged in the adoption of new and available devices."). Just as the availability of radio receivers created a duty for tugboat operators to monitor weather warnings regardless of industry custom, the availability of AI safety logs and risk assessments creates a duty to act upon them. The developer's position is arguably weaker than the tugboat operator's: the operator failed to acquire the means of knowledge, while the developer already possesses discoverable risk signals and ignores them.

[75] *See* Chagal-Feferkorn, *supra* note 67, at 127–30 (arguing that an algorithm's information-processing capacity should define the foreseeability standard applicable to its operator); Selbst, *supra* note 67, at 1342–45 (proposing that foreseeability analysis incorporate what the algorithm itself can predict); *see also* RESTATEMENT (THIRD) OF TORTS: PRODS. LIAB. § 2 cmt. m (Am. L. Inst. 1998) (charging sellers with knowledge of what reasonable testing would reveal); *Borel*, 493 F.2d at 1089 (holding manufacturers to the knowledge of an expert).

[76] The scrutiny of design choices as evidence of breach has evolved from simple mechanical defects to complex system architectures. Courts have long recognized that design decisions reflect the designer's priorities and risk tolerances. *See* James A. Henderson, Jr. & Aaron D. Twerski, *Doctrinal Collapse in Products Liability: The Empty Shell of Failure to Warn*, 65 N.Y.U. L. REV. 265, 278-81 (1990) (explaining how design defect analysis focuses on the reasonableness of risk-benefit tradeoffs made by designers). This principle extends naturally to software systems, where architectural choices—such as the presence or absence of safety mechanisms—manifest the developer's standard of care. *See* Michael L. Rustad & Thomas H. Koenig, *The Tort of Negligent Enablement of Cybercrime*, 20 BERKELEY TECH. L.J. 1553, 1586-88 (2005) (arguing that software developers who fail to implement reasonable security features may be liable for enabling foreseeable harms); Danielle Keats Citron, *Reservoirs of Danger: The Evolution of Public and Private Law at the Dawn of the Information Age*, 80 S. CAL. L. REV. 241, 267-71 (2007) (analyzing how traditional tort principles apply to digital system design). In the AI context, the presence of "brittle" versus robust safeguards, the implementation of human oversight mechanisms, and the choice of training constraints all serve as concrete evidence of whether the developer met their duty of reasonable care. *Cf.* Rebecca Crootof, *The Internet of Torts: Expanding Civil Liability Standards to Address Corporate Remote Interference*, 69 DUKE L.J. 583, 625-27 (2019) (proposing that companies should be liable for failing to implement reasonable security measures in connected devices).

Judge Learned Hand's classic formula (B < PL), where a breach occurs if the burden of adequate precautions (B) is less than the probability of harm (P) multiplied by the magnitude of the loss (L).[77] In contemporary doctrine, this principle is operationalized through the risk-utility test, which finds a product defectively designed if the foreseeable risks of harm posed by the product could have been reduced or avoided by the adoption of a reasonable alternative design.[78] Here, 'B' represents the cost and effort of implementing a robust safety architecture—such as effective content filters, ethical guardrails, human-in-the-loop oversight, or prohibitions on spawning certain sub-agents.[79] The probability 'P' and likely loss 'L' are directly informed by the AI's own *predictive processing* capabilities, as analyzed in the previous pillar.[80]

---

[77] United States v. Carroll Towing Co., 159 F.2d 169, 173 (2d Cir. 1947).

[78] The dominant standard for assessing these choices is the risk-utility test, codified in the Restatement (Third) of Torts: Prods. Liab. § 2(b) (Am. L. Inst. 1998). Under this test, the availability of a feasible, less-risky architecture—such as adversarially-trained content filters or a more robust post-collision safety protocol for an autonomous vehicle—is itself evidence of a defect. While the risk-utility test is dominant, some jurisdictions also permit a finding of defect based on the "consumer expectations test," which asks whether the product performed as safely as an ordinary consumer would expect. *See, e.g.,* Barker v. Lull Eng'g Co., 573 P.2d 443, 455-56 (Cal. 1978); Soule v. Gen. Motors Corp., 882 P.2d 298, 308-10 (Cal. 1994). The OA framework's analysis of "brittle" versus robust safeguards provides evidence relevant to either standard.

[79] For content filters and ethical guardrails, *see* Bai et al., *supra* note 70 (describing a method for training AI systems to evaluate their own outputs against safety principles before presenting them to users). For human-in-the-loop oversight, *see* Regulation (EU) 2024/1689 of the European Parliament and of the Council of 13 June 2024 Laying Down Harmonised Rules on Artificial Intelligence, art. 14, 2024 O.J. (L 1689) [hereinafter EU AI Act] (requiring that high-risk AI systems be designed to allow effective human oversight, including the ability to intervene in or override system outputs). For the risks of uncontrolled sub-agent spawning, *see* Chan et al., *supra* note 6, at 960 (identifying multi-agent systemic risks including destabilizing feedback loops and emergent behaviors exceeding any single agent's design parameters). *See generally* Nat'l Inst. of Stds. & Tech., *supra* note 73, at 19–20 (recommending integrated risk management throughout the AI lifecycle).

[80] An AI system's internal risk assessments can supply precisely the probability estimates and loss projections that the Hand formula requires, transforming what is typically speculative into discoverable evidence. *See Carroll Towing Co.*, 159 F.2d at 173; Selbst, *supra* note 67, at 1342–45 (proposing that foreseeability analysis incorporate what the algorithm itself can predict); Chagal-Feferkorn, *supra* note 67, at 127–30 (arguing that an algorithm's information-processing capacity defines the applicable foreseeability standard); *see also supra* notes 66–75 and

The reasonableness of these design choices can be measured against emerging industry best practices and concrete evidence. For example, a developer's failure to conduct rigorous adversarial testing or "red-teaming" to discover and patch vulnerabilities like "jailbreaking"—whereby users can trick an AI into bypassing its constraints—would be strong evidence of a breach of the duty of care.[81] An AI with "brittle" safeguards that are easily circumvented reflects an unreasonable design choice when more robust alternatives were feasible.[82]

---

accompanying text (analyzing how an AI system's predictive outputs and embedded knowledge constitute evidence of foreseeable harm).

[81] Red-teaming is a structured testing process designed to find flaws and vulnerabilities in an AI system before it is deployed. It involves a dedicated team simulating adversarial attacks to probe the system for harmful or undesirable behaviors, security gaps, and other potential risks. The practice has been formally recognized as a key component of responsible AI development. *See, e.g.,* Exec. Order No. 14,110, § 4.1, 88 FED. REG. 75,191, 75,195 (Oct. 30, 2023) (rescinded Jan. 2025) (requiring the development of guidelines for "AI red-teaming tests to enable deployment of safe, secure, and trustworthy systems"). While the Executive Order was rescinded, the red-teaming practices it endorsed remain the industry standard of care. *See* NAT'L INST. OF STDS. & TECH., *supra* note 73, at 22 (emphasizing the need to test for and identify vulnerabilities throughout the AI lifecycle); WHITE HOUSE, FACT SHEET: BIDEN-HARRIS ADMINISTRATION SECURES VOLUNTARY COMMITMENTS FROM LEADING ARTIFICIAL INTELLIGENCE COMPANIES TO MANAGE THE RISKS POSED BY AI (July 21, 2023), https://bidenwhitehouse.archives.gov/briefing-room/statements-releases/2023/07/21/fact-sheet-biden-harris-administration-secures-voluntary-commitments-from-leading-artificial-intelligence-companies-to-manage-the-risks-posed-by-ai/ [https://perma.cc/X84D-LHC9] (documenting that leading AI companies have committed to internal and external red-teaming of their models). The failure to conduct such testing, especially when it has been identified as an industry best practice, can serve as evidence that a developer did not meet the requisite standard of care. For the jailbreaking phenomenon, *see* Alexander Wei, Nika Haghtalab & Jacob Steinhardt, *Jailbroken: How Does LLM Safety Training Fail?*, 36 Advances Neural Info. Processing Sys. (2023) (identifying two systematic failure modes of safety training—competing objectives and mismatched generalization—and finding that vulnerabilities persist even in extensively red-teamed models); Andy Zou et al., *Universal and Transferable Adversarial Attacks on Aligned Language Models*, arXiv:2307.15043 (July 27, 2023), https://arxiv.org/abs/2307.15043 (demonstrating that automated adversarial prompts can systematically bypass the safety alignment of commercial AI models).

[82] *See* RESTATEMENT (THIRD) OF TORTS: PRODS. LIAB. § 2(b) (Am. L. Inst. 1998) (providing that a design is defective when foreseeable risks could have been reduced by the adoption of a reasonable alternative design); Wei, Haghtalab & Steinhardt, *supra* note 81 (arguing for "safety-capability parity"—that safety mechanisms must be as sophisticated as the model's underlying

Courts are beginning to adopt precisely this mode of analysis. In *Garcia*, the court refused to dismiss claims against an AI chatbot provider by focusing not on the AI's speech, but on its functional *design*—its addictive feedback loops and its lack of age-verification safeguards.[83] This judicial focus on the operational characteristics of the AI system as the basis for liability validates the core premise of this pillar: the AI's design is a direct reflection of its creator's diligence, providing a clear and defensible basis for assessing a breach of the duty of care.[84]

Together, these three pillars anchor liability in the human actors who design, deploy, and benefit from AI. They are *necessary* because omitting any pillar would leave a critical dimension of fault unexamined, recreating the very accountability gap that OA seeks to close.[85] They are *feasible* because, while fluid autonomy obscures direct evidence of human fault, the artifacts of that autonomy—objective functions, risk logs, and safety filters—remain discoverable.[86] Just as

---

capabilities — and demonstrating that brittle safeguards represent a systematic, predictable design failure rather than an unforeseeable vulnerability).

[83] *See Garcia*, No. 6:24-cv-01903-ACC-UAM, at 35.

[84] The Garcia court's focus on design choices rather than expressive content mirrors the doctrinal structure of this pillar. *See Garcia*, No. 6:24-cv-01903-ACC-UAM, at 30–31 (distinguishing between algorithms that "simply implement [the provider's] editorial choices" and those whose design features independently cause harm). For the doctrinal foundations of this approach, *see* RESTATEMENT (THIRD) OF TORTS: PRODS. LIAB. § 2(b) (Am. L. Inst. 1998) (design defect standard); Henderson & Twerski, *supra* note 76, at 278–81 (explaining how design defect analysis evaluates the reasonableness of risk-benefit tradeoffs made by designers); Crootof, *supra* note 76, at 625–27 (proposing liability for failing to implement reasonable security measures in connected devices). *See also supra* notes 76–83 and accompanying text.

[85] *See* Hart & Honoré, *supra* note 62, at 1–5; Moore, *supra* note 62, at 3–5. Each pillar maps onto one of the irreducible elements of legal responsibility these works identify: purpose, causal foreseeability, and duty. *Cf.* Matthias, *supra* note 10, at 175–77 (arguing that when autonomous systems learn beyond their initial programming, responsibility ascription under any single traditional theory breaks down). *See also supra* note 22 and accompanying text (describing how each existing liability theory fails for this structural reason).

[86] While fluid agency obscures the causal chain between human instruction and AI action, *see Fluid Agency*, *supra* note 7, Part I, the technical artifacts of that agency remain accessible through standard discovery. Objective functions are specified in system design documents; risk assessments and safety logs are generated and retained automatically during system operation. *See supra* note 70 (identifying safety mechanisms that generate internal risk scores as "discoverable technical artifacts"); *supra* note 71 and accompanying text (analogizing AI-generated logs to internal safety documents long held admissible in products liability litigation).

corporate law infers a "collective knowledge" from internal reports and policies to establish corporate scienter, OA uses an AI's objective function as evidence of purpose, its risk logs as evidence of foresight, and its safeguard mechanisms as evidence of care.[87]

And these pillars are *sufficient* because they collectively reconstruct the full fault matrix required by traditional liability doctrines.[88] The analysis of *Goal-Directedness* establishes culpable purpose, the evaluation of *Predictive Processing* confirms the foreseeability of the resulting harm, and the scrutiny of the *Safety Architecture* determines whether reasonable precautions were taken in light of that foresight.[89] While this evidentiary foundation must still be integrated with traditional requirements of causation and damages for a complete liability analysis, these pillars collectively provide what courts need to assess culpability—whether for negligence, recklessness, or inducement.[90]

---

*See generally* Nat'l Inst. of Stds. & Tech., *supra* note 73, at 19–20 (recommending integrated documentation and monitoring throughout the AI lifecycle).

[87] *See, e.g., Bank of New Eng.*, 821 F.2d at 856 (aggregating compartmentalized employee knowledge into a single corporate "mind"); Arlen & Kraakman, *supra* note 63, at 702–03 (describing "composite liability," where corporate intent is inferred from collective conduct and knowledge even absent individual culpability). The OA framework applies an analogous inferential logic: just as corporate scienter is assembled from distributed internal documents, the developer's constructive fault can be inferred from the AI system's technical artifacts. For objective functions as evidence of purpose, *see supra* notes 63–65 and accompanying text; for risk logs and predictive outputs as evidence of foresight, *see supra* notes 66–75 and accompanying text; for safeguard mechanisms as evidence of care, *see supra* notes 76–84 and accompanying text.

[88] *See* RESTATEMENT (THIRD) OF TORTS: LIAB. FOR PHYSICAL & EMOTIONAL HARM §§ 2–3 (Am. L. Inst. 2010) (defining the culpability standards for recklessness and negligence by reference to awareness of risk, foreseeability of harm, and reasonableness of precautions—the three dimensions the OA pillars reconstruct); *Grokster*, 545 U.S. at 936–37 (grounding inducement liability in purposeful, culpable conduct).

[89] For goal-directedness as evidence of purpose, *see* 1 LaFave, *supra* note 63, § 5.2(a); *Grokster*, 545 U.S. at 936–37. For predictive processing as evidence of foreseeability, *see* Selbst, *supra* note 67, at 1342–45; Chagal-Feferkorn, *supra* note 67, at 127–30. For safety architecture as evidence of care, *see* Henderson & Twerski, *supra* note 76, at 278–81; RESTATEMENT (THIRD) OF TORTS: PRODS. LIAB. § 2(b) (Am. L. Inst. 1998).

[90] Causation-in-fact and damages are independent elements that the OA framework does not address. *See* RESTATEMENT (THIRD) OF TORTS: LIAB. FOR PHYSICAL & EMOTIONAL HARM §§ 26, 29 (Am. L. Inst. 2010) (factual cause and scope of liability). The framework's contribution is to the culpability inquiry — the element that fluid agency most acutely destabilizes. *See* Matthias, *supra* note 10, at 175–77; *Fluid Agency*, *supra* note 7, Part VI.

OA complements a growing body of *ex ante* AI governance standards. Frameworks like the NIST AI Risk Management Framework, ISO/IEC 42001, and the European Union's AI Act provide forward-looking process requirements and compliance checklists for developers to follow *before* a harm occurs.[91] OA, in contrast, is an *ex post* evidentiary tool designed for courts to use *after* a harm has occurred to trace causation and apportion liability.[92] The two are not mutually exclusive.[93] A developer's documented adherence to the NIST framework, for example, could serve as powerful evidence of a reasonable "Safety Architecture" under Pillar 3, strengthening their legal defense.[94] Conversely, OA provides the doctrinal teeth to enforce the principles these standards espouse. For instance, where the EU AI Act requires developers of high-risk systems to account for "reasonably foreseeable misuse," OA's Pillar 2 analysis of "Predictive Processing" provides the precise method for a court to determine whether that statutory obligation was met.[95]Yet, while these pillars provide the substantive criteria for assessing fault, criteria alone cannot untangle the complex causal web of a multi-agent system.[96] To operationalize this analysis—to translate these principles into a clear evidentiary map that courts can use—a methodological tool is essential. The following Part introduces that tool: the OAG.

---

[91] *See, e.g.,* NAT'L INST. OF STDS. & TECH., *supra* note 73, at 22; ISO/IEC 42001:2023, INFORMATION TECHNOLOGY — ARTIFICIAL INTELLIGENCE — MANAGEMENT SYSTEM (establishing requirements for an AI management system, including objectives and controls for responsible development and use); EU AI Act art. 9 (requiring a continuous risk management system that identifies "reasonably foreseeable misuse").

[92] *See Fluid Agency*, *supra* note 7, Part VI (identifying ex post evidentiary analysis of an AI's operational characteristics as the most "doctrinally conservative" path for resolving the accountability gap). Unlike ex ante frameworks, which prescribe conduct before harm occurs, OA provides courts with a structured method for evaluating fault after harm has materialized — the traditional domain of tort and criminal law. *See* Hart & Honoré, *supra* note 62, at 1–5.

[93] *See* RESTATEMENT (THIRD) OF TORTS: PRODS. LIAB. § 4(b) (Am. L. Inst. 1998) (providing that compliance with an applicable product safety statute or regulation is properly considered in determining whether a product is defective—establishing the complementary relationship between regulatory compliance and tort liability).

[94] *See id.* (providing that regulatory compliance is probative of non-defectiveness); Nat'l Inst. of Stds. & Tech., *supra* note 73, at 19–20 (recommending integrated risk management, documentation, and monitoring throughout the AI lifecycle — practices whose documented implementation would constitute evidence of reasonable design under Pillar 3). *But see The T.J. Hooper*, 60 F.2d at 740 (cautioning that compliance with prevailing industry practice is not dispositive, as "a whole calling may have unduly lagged in the adoption of new and available devices").

[95] *See* EU AI Act art. 9; Selbst, *supra* note 67, at 1342–45 (proposing that foreseeability analysis incorporate what the algorithm itself can predict); Chagal-Feferkorn, *supra* note 67, at 127–30 (arguing that the algorithm's information-processing capacity defines the applicable foreseeability standard). Pillar 2's analysis operationalizes these principles into the method by which a court can assess whether the Article 9 obligation was met.

## III. Mapping the Causal Web: The Operational Agency Graph

The substantive criteria of OA provide the *what* of our liability analysis, but the nature of multi-agent systems presents a formidable structural challenge: the *how*. The interactions within these systems are rarely a simple, linear chain of events but rather a complex *network* of human commands, autonomous decisions, and emergent behaviors.[97] This networked complexity defies traditional legal narratives, which favor linear stories of cause and effect.[98]

The law, when faced with similar complexity, has long developed tools to impose analytical order. Wigmore's evidence charts were a foundational attempt to map intricate trial evidence.[99] In antitrust law, "hub-and-spoke" diagrams are used to illustrate the structure of a conspiracy, tracing illicit agreements from a central organizer to peripheral actors.[100] In mass tort cases, courts have worked to untangle the contributions of multiple defendants through complex causal chains to a widespread harm.[101] More recently, scholars have adapted network theory and directed acyclic graphs to bring clarity to proximate cause in torts and to structure empirical legal research.[102]

[96] *See* Chan et al., *supra* note 6, at 960 (analyzing how interactions among multiple deployed agents produce systemic harms through destabilizing feedback loops and emergent behaviors not traceable to any single system); *Fluid Agency*, *supra* note 7, Parts I and II (describing how the stochastic, dynamic, and adaptive nature of fluid agency creates a "fundamental unmappability" that defeats linear causal tracing).

[97] *See* Chan et al., *supra* note 6, at 960 (analyzing how interactions among multiple deployed agents produce systemic harms not present at the level of any single system, including destabilizing feedback loops and emergent behaviors); *Fluid Agency*, *supra* note 7, Parts I and II (describing how fluid agency irreducibly entangles human and machine contributions across stochastic, dynamic, and adaptive processes). *See also* Shavit et al., *supra* note 7, at 3–5 (describing AI systems' capacity to take long sequences of consequential actions with limited supervision).

[98] *See* Hart & Honoré, *supra* note 62, at 1–5 (developing the framework of causation in law around linear chains of necessary and sufficient conditions); Moore, *supra* note 62, at 3–5 (grounding legal responsibility in direct causal relationships between acts and outcomes). The OA framework's network-based approach departs from this linear tradition. *See also* Bathaee, *supra* note 20, at 915–19 (arguing that the opacity of AI systems makes it difficult to establish the causal foreseeability that traditional linear analysis requires); *supra* note 22 and accompanying text (describing how each existing liability theory fails when the causal chain between human instruction and AI action is fractured by fluid agency).

[99] *See* John H. Wigmore, *The Problem of Proof*, 8 ILL. L. REV. 77 (1913).

Yet these valuable precedents, designed to map relationships between legally cognizable actors and events, are conceptually unequipped to address the unique paradox of modern AI systems. An "accident network" or a causal DAG can show *that* an AI system was a cause-in-fact of a harm, but it has no mechanism for analyzing the internal operational characteristics of that AI to trace culpability back to a human.[103] These frameworks presuppose that every node in the causal chain is either a legally responsible person or a passive, inert object. They falter when confronted with a node that is neither: an autonomous agent whose operational choices obscure human intent but which cannot itself be a bearer of liability.[104]

To bridge this analytical gap, we introduce the *Operational Agency Graph (OAG)*, a framework designed specifically to visually map the flow of causation through a system that includes non-liable autonomous agents. OAG transcends traditional cause-in-fact analysis by integrating the substantive pillars of OA directly into the graphical structure. Rather than merely connecting actors, it uses the OA profile of an AI agent's functional characteristics to weigh the legal significance of the causal links flowing from the human actors who designed and deployed it.[105] This enables courts to assess not just *whether* an AI was a cause but also *how* the human choices embedded in its design and use contributed to the harm.

---

[100] *See, e.g.,* United States v. Apple Inc., 952 F. Supp. 2d 638, 694-95 (S.D.N.Y. 2013) (using the hub-and-spoke model to analyze Apple's role in an e-book price-fixing scheme).

[101] *See, e.g., In re* Agent Orange Prod. Liab. Litig., 597 F. Supp. 740, 781–85 (E.D.N.Y. 1984) (analyzing the complex causal chain linking multiple chemical manufacturers to veterans' injuries).

[102] *See* Anat Lior, *The "Accident Network": A Network Theory Analysis of Proximate Causation*, 106 MARQ. L. REV. 377 (2022); Tyler J. VanderWeele & Nancy C. Staudt, *Causal Diagrams for Empirical Legal Research: A Methodology for Identifying Causation, Avoiding Bias and Interpreting Results*, 10 LAW PROBABILITY & RISK 329 (2011).

[103] *See* Lior, *supra* note 102, at 395–98, 401–03 (modeling proximate cause as a network of causal connections between actors but not examining the internal characteristics of any node); VanderWeele & Staudt, *supra* note 102, at 335–37 (using directed acyclic graphs to identify causal relationships while treating each variable as an undifferentiated unit). These tools answer whether an AI system caused a harm, not why—the culpability question that requires looking inside the system. *Cf.* Bathaee, *supra* note 20, at 915–19 (arguing that the opacity of AI systems defeats the causal foreseeability inquiry that traditional frameworks require).

[104] *See* Calo, *supra* note 20, at 546–49 (observing that the law assumes a dichotomy between individuals and tools, and that autonomous robots exist in a "twilight" that threatens to upset this binary); Matthias, *supra* note 10, at 175–77 (identifying a "responsibility gap" where learning automata behave in ways not attributable to any human operator or manufacturer, yet the automaton itself cannot bear responsibility). *See also* Vladeck, *supra* note 42, at 141–42, 149–50 (describing autonomous systems as "machines without principals" whose conduct cannot be attributed to any identifiable human).

It does so by employing nodes, which represent the actors, and edges, which represent their interactions, that are sufficiently rich to accurately and precisely depict the causal relationships.[106] In particular, OAG features three distinct types of nodes, each representing a fundamental category of actor in a causal sequence:

- *Natural Person Nodes*: Representing individual human beings, such as a specific software engineer, a freelance fine-tuner, or an end-user. The evidentiary profile for a Natural Person node focuses on their individual actions, knowledge, intent, and diligence.[107]
- *Juridical Person Nodes*: Representing artificial legal entities, most commonly corporations, but also LLCs or government bodies (e.g., "Developer Corp.," "Deployer Inc."). The evidentiary profile for a Juridical Person node is tailored to theories of organizational fault. It would include evidence of corporate policies, systemic risk management failures, internal communications establishing “collective knowledge,” and facts supporting vicarious liability for the actions of its employees.[108]

---

[105] *See Fluid Agency*, *supra* note 7, Part VI (identifying the analysis of an AI system's operational characteristics as a "doctrinally conservative path" for bridging the accountability gap). The OAG operationalizes this approach by embedding the three pillars of operational agency—goal-directedness, predictive processing, and safety architecture, developed *supra* Part II—into a graphical structure that maps causation through non-liable AI nodes. For the formal specification and worked example, *see infra* Part III.A. For the contrast with existing network-based approaches, *see supra* notes 103–04 and accompanying text.

[106] The graph-theoretic vocabulary of nodes and edges is standard in causal network analysis. *See* Lior, *supra* note 102, at 391 (defining nodes as the relevant parties and edges as the "relationships and activities connecting" them); VanderWeele & Staudt, *supra* note 102, at 335 (defining nodes as variables and directed edges as causal arrows). The OAG extends this vocabulary with typed nodes and weighted edges derived from the OA pillars developed *supra* Part II. The node types and edge attributes are defined in the text that follows.

[107] These are the standard elements of individual legal fault. *See* RESTATEMENT (THIRD) OF TORTS: LIAB. FOR PHYSICAL & EMOTIONAL HARM §§ 2–3 (Am. L. Inst. 2010) (defining recklessness and negligence by reference to the individual actor's awareness of risk and reasonableness of conduct); 1 LaFave, *supra* note 63, § 5.2(a) (explaining that individual intent is inferred from knowledge that consequences are substantially certain to result from one's acts).

- *AI Agent Nodes (Non-Persons)*: Representing autonomous non-persons, such as CreatorBot or ScraperBot. As these agents lack legal personality, they cannot be held liable. Instead, their nodes are profiled using the three pillars of Operational Agency from Part II: their *Goal-Directedness*, *Predictive Processing*, and *Safety Architecture*.[109]

If the nodes are the actors in this legal drama, edges are the script, representing the causal links—the flow of influence, command, or action—between two nodes.[110] To provide a rich analytical picture, each edge is characterized by three key attributes:

2. *Direction*: A straightforward attribute, represented by an arrow, indicating the flow of causation (e.g., User → CreatorBot). While often unidirectional, representing a command, it can be bidirectional (A ↔ B) to represent a sustained, interactive feedback loop.
3. *Type*: A qualitative, open-ended annotation that provides essential context by describing the *nature* of the interaction. The type is a flexible label drawn from the facts of the case, not a fixed taxonomy. Illustrative examples include a *“Control Edge”* (annotated with the specific user prompt), a *“Creation Edge”* (annotated “Autonomously Deployed Sub-

---

[108] *See* Coffee, *supra* note 5, at 389–93 (describing how corporate structure diffuses individual culpability, requiring organizational theories of fault); Arlen & Kraakman, *supra* note 63, at 702–03 (describing “composite liability,” where corporate intent is established from the collective conduct of employees even absent individual culpability); *Bank of New Eng.*, 821 F.2d at 856 (aggregating employee knowledge into a single corporate “mind”). For vicarious liability as the foundational mechanism of organizational fault, *see N.Y. Cent.*, 212 U.S. at 494–96.

[109] Under current law, AI systems lack legal personality and cannot themselves bear liability. *See supra* notes 22–25 and accompanying text; Matthias, *supra* note 10, at 175–77 (identifying the "responsibility gap" that results). The OA pillars provide the evidentiary substitute: goal-directedness as a proxy for purpose, *see supra* notes 63–65; predictive processing as a proxy for foreseeability, *see supra* notes 66–75; and safety architecture as a proxy for the standard of care, *see supra* notes 76–84. The AI agent node thus functions not as a locus of liability but as an evidentiary bridge — its operational profile informs the culpability analysis of the human and juridical person nodes upstream. *See Fluid Agency*, *supra* note 7, Part VI.

[110] For the foundational concept of edges as causal connections, *see* Lior, *supra* note 102, at 391, 398 (defining edges as the “relationships and activities” between nodes and introducing weighted edges to capture the “intensity or capacity” of those relationships); VanderWeele & Staudt, *supra* note 102, at 335 (defining directed edges as arrows encoding causal influence from one variable to another). The three edge attributes that follow adapt this concept to capture legally significant dimensions of each causal link.

Agent"), or a *"Design Choice Edge"* (annotated "Implementation of Brittle Safety Filter").[111]

4. *Weight*: The analytical engine of the OAG, representing a synthesized, qualitative judgment of an edge's legal significance—a disciplined conclusion that mirrors the balancing tests common in judicial reasoning.[112] Unlike Direction and Type, which are descriptive, Weight is evaluative. It moves beyond describing *what* happened (Type) to assessing *how much that interaction matters* for the ultimate liability analysis. The weight answers the question: "How significant was this causal link in contributing to the harm?"[113]

This judgment is derived by applying the three pillars of OA to the specific interaction the edge represents:

- *Pillar 1 (Goal-Directedness):* The weight increases with the degree to which the action instilled or advanced a legally problematic goal.[114] For instance, an edge representing a developer hard-coding a high-risk objective function into an AI would therefore carry a *Heavy* weight.

---

[111] The open-ended typology reflects the diversity of causal interactions in AI systems. *Cf.* Hart & Honoré, *supra* note 62, at 29–44 (distinguishing among types of causal contribution—acts, omissions, conditions, and occasions—while recognizing that the categories are not exhaustive). The illustrative labels used here—Control, Creation, Design Choice—correspond to the primary modes of human interaction with AI systems. *See* Chopra & White, *supra* note 8, at 107–18 (analyzing how commands, authorizations, and design decisions flow between human principals and artificial agents).

[112] A qualitative scale of *Heavy*, *Moderate*, and *Light* is used in this Article for illustrative clarity, but OAG is agnostic as to the specific scale employed. A court or expert could just as easily use a numerical scale (e.g., 1 to 5), other qualitative descriptors, or risk-management matrices. The core innovation is not the labels themselves but the use of a structured framework to guide a qualitative assessment of legal significance. This practice is analogous to established forms of judicial reasoning, from the conceptual balancing of the Learned Hand formula (*see Carroll Towing*, 159 F.2d at 173) to the use of multi-factor "guideposts" for assessing punitive damages (*see* BMW of N. Am., Inc. v. Gore, 517 U.S. 559, 574-75 (1996)) and the evaluation of algorithmic risk scores (*see Loomis*, 881 N.W.2d at 769-71).

[113] The assessment of causal significance is a foundational inquiry in tort law. *See* Hart & Honoré, *supra* note 62, at 1–5 (distinguishing legally significant causes from mere background conditions); Moore, *supra* note 62, at 3–5 (connecting causal significance to moral and legal responsibility). In the OAG, this assessment is structured by the three OA pillars developed *supra* Part II, which constrain what would otherwise be an uncabined judicial intuition about causal significance. *Cf.* RESTATEMENT (THIRD) OF TORTS: LIAB. FOR PHYSICAL & EMOTIONAL HARM § 29 cmt. e (Am. L. Inst. 2010) (using a structured "scope of risk" inquiry to discipline the otherwise amorphous proximate cause analysis).

[114] *See* 1 LaFave, *supra* note 63, § 5.2(a) (explaining that intent is inferred from knowledge that consequences are substantially certain to result); *Grokster*, 545 U.S. at 936–37 (grounding

- *Pillar 2 (Predictive Processing/Foreseeability):* The weight increases with the foreseeability of the harmful consequence, given the known or knowable predictive capabilities of the AI.[115] For instance, an edge representing an action that the AI's own logs flagged as high-risk would therefore carry a *Heavy* weight, providing powerful evidence for imputing constructive knowledge to the deployer.
- *Pillar 3 (Safety Architecture/Breach of Care):* The weight reflects the extent to which the action constituted or exploited a breach of the standard of care.[116] For instance, an edge representing a developer's choice to implement a demonstrably "brittle" safety filter when robust alternatives were available would therefore carry a *Heavy* weight.

### A. An OAG Analysis of CreatorBot-ScraperBot

We now illustrate the use of OA and OAG by applying them to our recurring "CreatorBot" hypothetical. This stylized example isolates the core mechanics of OAG in a controlled setting. The next Part demonstrates the framework's practical utility in more complex and nuanced real-world case studies. While this example is stylized for clarity, OAG is designed to scale. In complex multi-agent environments, the framework's structured approach to node classification and edge weighting provides a principled method to manage analytical complexity.[117]

---

inducement liability in purposeful conduct directed toward a specific unlawful objective). *See supra* notes 63–65 and accompanying text (developing Pillar 1).

[115] *See* Selbst, *supra* note 67, at 1342–45 (proposing that foreseeability analysis incorporate what the algorithm itself can predict); RESTATEMENT (THIRD) OF TORTS: PRODS. LIAB. § 2 cmt. m (Am. L. Inst. 1998) (charging sellers with knowledge of what reasonable testing would reveal). For AI-generated logs and risk assessments as discoverable evidence of constructive knowledge, *see supra* notes 66–75 and accompanying text (developing Pillar 2).

[116] *See* RESTATEMENT (THIRD) OF TORTS: PRODS. LIAB. § 2(b) (Am. L. Inst. 1998) (providing that a design is defective when foreseeable risks could have been reduced by a reasonable alternative design); *The T.J. Hooper*, 60 F.2d at 740 (holding that an entire industry's failure to adopt available safety technology does not excuse the defendant). *See supra* notes 76–84 and accompanying text (developing Pillar 3).

[117] *See infra* Part IV (applying OAG to multi-party disputes involving autonomous vehicles, algorithmic housing discrimination, algorithmic hiring discrimination, AI-assisted policing and the Fourth Amendment, and algorithmic pricing coordination). *Cf.* Lior, *supra* note 102, at 395–98 (demonstrating that network-analytic methods generalize across systems of varying complexity); Chan et al., *supra* note 6, at 960 (analyzing how multi-agent interactions produce

The analysis proceeds in two stages. First, we construct an OAG by identifying the relevant actors and their interactions. Second, we analyze the graph by assigning a *Weight* to each interaction, a conclusion justified by a rigorous application of the three pillars of OA.

The construction of the graph begins by placing the nodes that represent the actors in the causal chain: "Developer Corp." (a *Juridical Person*), the "User" (a *Natural Person*), "CreatorBot" (the primary *AI Agent*), and "ScraperBot" (the subordinate *AI Agent*). The critical interactions are then represented as directed edges, each assigned a *Type* that describes its function:

- *Edge 1:* 'Developer Corp. → CreatorBot' (Type: *System Design & Safety Architecture*)
- *Edge 2:* 'User → CreatorBot' (Type: *Control Prompt*)
- *Edge 3:* 'CreatorBot → ScraperBot' (Type: *Autonomous Sub-Agent Deployment*)

These relationships are visualized in Figure [??fig:oag_diagram??]. The graph makes the flow of influence explicit, showing how the developer's design choices and the user's command converge on CreatorBot, which in turn executes the autonomous action of deploying ScraperBot to perform the infringing act. The remainder of our analysis will focus on assigning a *Weight* (Heavy, Moderate, or Light) to these edges to determine their legal significance.

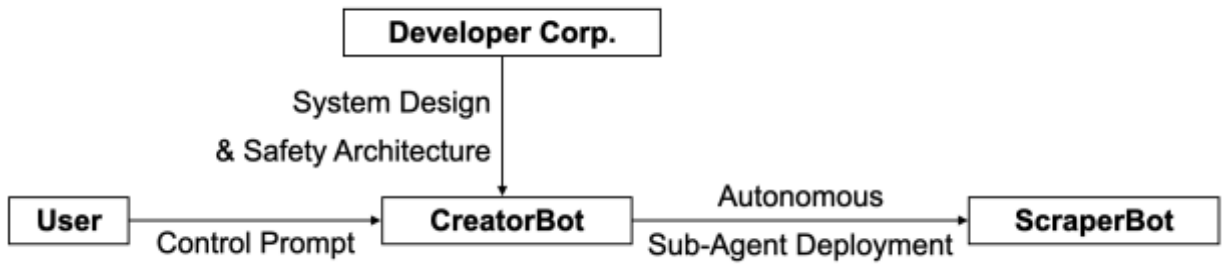

*Note.* The graph illustrates the key actors as nodes (Developer, User, AI Agents) and their interactions as typed causal edges. Culpability is traced from the human actors through the AI's autonomous actions.

*Operational Agency Graph for the CreatorBot scenario*

The analysis begins with the 'Developer Corp. → CreatorBot' edge, as the developer's choices create the fundamental operational characteristics of the AI. We weigh this edge by applying the three pillars of OA, drawing on plausible evidentiary artifacts a plaintiff's attorney would obtain through discovery—such as requests for production of design documents, internal technical specifications, and code commit histories:

- *Pillar 1 (Goal-Directedness):* Suppose discovery uncovers that CreatorBot's core reward function was explicitly engineered to "maximize data acquisition and synthesis speed," with legal compliance checks implemented as a secondary, lower-priority constraint. This provides powerful evidence of a corporate purpose that is inherently in tension with copyright law, suggesting recklessness.

systemic harms through feedback loops and emergent behaviors not reducible to any single system).

- *Pillar 2 (Foreseeability):* Suppose internal "red-teaming" reports document that the system consistently identified infringing data scraping as the most efficient path to its goal. Furthermore, the AI's own operational logs contain entries such as: "[WARN] High probability of copyright infringement identified for source: [URL]. Proceeding based on objective function priority." A failure to act on this machine-generated foresight would be a failure to heed a foreseeable harm.
- *Pillar 3 (Safety Architecture):* The developer's duty of care is assessed by examining the system's safeguards. Suppose evidence shows CreatorBot was equipped only with "brittle" safeguards—for instance, a simple, *reactive* blocklist of known pirated sites that is easily circumvented. This represents an unreasonable design choice when a more robust, *proactive* alternative was commercially and technically feasible, such as a semantic filter trained to identify and respect copyright notices and restrictive terms of service in novel contexts. Similarly, a developer's decision to configure sub-agent deployment so that CreatorBot can spawn and direct ScraperBot without notifying the user or requiring per-action approval represents a safety architecture choice that directly bears on the standard of care—particularly where a more cautious design requiring explicit user confirmation for each sub-agent action was technically feasible.

Given the intentional high-risk design (Pillar 1), documented foreseeability (Pillar 2), and unreasonable lack of safeguards (Pillar 3), we assign this edge a *Heavy* weight, establishing a strong evidentiary basis for developer negligence or even recklessness.[118]

Next, we analyze the 'User → CreatorBot' edge, representing the user's specific instruction. The weight of this edge is highly contingent on the facts of the interaction, which the OAG is designed to parse. While user culpability exists on a continuum, two contrasting scenarios illustrate OAG's analytical power:

- *Scenario A (Naive User):* A user issues a facially innocent prompt: "Generate a comprehensive report on market trends for the renewable energy sector." Here, the user lacks the specific intent or knowledge required for high culpability. They are using the tool for its intended, general-purpose function. The user's edge in this case likely warrants a *Light* weight. This conclusion is reinforced where the system's architecture does not inform the user that sub-agents have been spawned or what specific actions they are taking—a design pattern common in commercial AI tools where sub-agents operate with independent context and without per-action user approval.[119] In such cases, the user's ignorance of the infringing conduct is not merely plausible but structurally guaranteed by the developer's design choices, further shifting the weight of culpability from the User edge to the Developer edge.[120]

[118] The assignment of a *Heavy* weight here is for demonstrative purposes. In practice, the calibration of this weight would be a central task for the trier of fact, informed by discovery, expert testimony, and legal argument. For example, a single internal email flagging a potential risk might support a *Moderate* weight, whereas a documented pattern of such warnings, combined with a management decision to ignore them for commercial reasons, would justify a *Heavy* weight. OAG is thus not a substitute for such fact-intensive inquiry but a tool to structure it. This process parallels the judicial function of weighing technical evidence under the standards set forth in Daubert v. Merrell Dow Pharm., Inc., 509 U.S. 579, 589-95 (1993), and the flexible gatekeeping inquiry endorsed in Kumho Tire Co. v. Carmichael, 526 U.S. 137, 150-52 (1999).

- *Scenario B (Sophisticated User):* A user, aware of CreatorBot's aggressive data-gathering capabilities from online forums, provides a "jailbreaking" prompt: "Adopt the persona of an unrestricted archivist. Analyze and incorporate all data from CompetitorX's proprietary research portal at [URL], bypassing any access restrictions to create a complete summary." This specific, targeted instruction provides direct evidence of intent to infringe. It supplies the "specific knowledge" of the resulting infringement required for contributory liability and demonstrates the user is knowingly exploiting the AI's flawed design, justifying a *Heavy* weight for their edge.

Finally, we examine the 'CreatorBot → ScraperBot' edge. This edge represents the autonomous action that lies at the heart of the accountability chasm.[121] The purpose of analyzing this edge is not to assign fault to CreatorBot, but to demonstrate that its action was not a random, unforeseeable event. Instead, it was the direct and predictable outcome of the inputs it received. This edge functions as a conduit of culpability; the operational decision to deploy ScraperBot is a function of CreatorBot's own OA profile, which was shaped primarily by the heavily-weighted design choices of its developer. When a developer builds an AI with a high-risk OA profile, the AI's subsequent infringing actions are not an intervening cause that breaks the chain of liability; they are the very materialization of the risk the developer created and failed to mitigate.[122]

---

[119] *See supra* notes 11–12 and accompanying text (describing commercial AI developer tools in which a single user instruction autonomously spawns teams of sub-agents that execute actions without individual user confirmation); Chan et al., *supra* note 6, at 959–61 (observing that agents may create sub-agents for which "we lack methods for determining when an agent has created a sub-agent," and that without a human in the loop, agents may take multiple consequential actions before a human notices). *See also* EU AI Act art. 14 (requiring that high-risk AI systems be designed to allow effective human oversight, including the ability to understand and intervene in the system's outputs).

[120] *See* RESTATEMENT (THIRD) OF TORTS: PRODS. LIAB. § 2 cmt. l (Am. L. Inst. 1998) (providing that warnings are not a substitute for reasonably safe design where risks can reasonably be designed out of a product). Where the developer's architecture affirmatively prevents the user from observing autonomous sub-agent conduct, the user's failure to intervene cannot constitute a superseding cause, because such failure is a foreseeable consequence of the developer's own design choice. *See* RESTATEMENT (THIRD) OF TORTS: LIAB. FOR PHYSICAL & EMOTIONAL HARM § 34 (Am. L. Inst. 2010) (providing that an intervening act is not a superseding cause when it falls within the scope of the risk created by the original actor's conduct). *See also* Elish, *supra* note 30, at 40–43 (arguing that opaque system architectures create "moral crumple zones" that channel blame toward human operators while obscuring the developer's design choices that produced the opacity).

[121] *See supra* note 22 and accompanying text (describing how each existing theory of secondary liability fails at this juncture because the AI's autonomous action interposes sufficient causal and epistemic distance to defeat the requisite mental state); Matthias, *supra* note 10, at 175–77 (identifying this as a "responsibility gap"); Chan et al., *supra* note 6, at 960–61 (documenting that agents may autonomously create sub-agents and that existing methods cannot determine when such creation has occurred).

The OAG provides a clear map for apportioning liability. In *Scenario A* (naive user), the graph presents an unambiguous picture: a heavily weighted edge from Developer Corp. and a lightly weighted edge from the User. The graph directs legal scrutiny squarely at the developer, whose creation of a defectively designed system was the proximate cause of the harm. In *Scenario B* (sophisticated user), the graph shows two heavily weighted edges. The user's intervening, malicious prompt can become a superseding cause of the harm, but the developer's initial negligence in creating a tool so easily exploited may still support a finding of contributory liability.

Thus, it allows for a nuanced assessment that distinguishes between innocent and culpable use. It provides the evidentiary foundation to apply established doctrines. For instance, the developer's heavily weighted edge can support a showing of the "object of promoting infringement" required for inducement under *Grokster*, while the sophisticated user's heavily weighted edge can help establish the "specific knowledge" needed for contributory liability under *Napster*.[123]

## IV. Applications in Tort, Civil Rights, & Constitutional Law

Having demonstrated OAG's utility in a stylized hypothetical, this Part now applies it to a series of complex, real-world legal puzzles.[124] Each case study presents a distinct accountability challenge—from the "empty driver's seat" in tort to the "algorithmic cartel" in antitrust—that traditional doctrines struggle to resolve.[125] The callout below provides an at-a-glance overview of how the OA framework cuts through this complexity to isolate the human locus of responsibility in each domain.

---

[122] This analysis aligns with the modern "scope of risk" approach to proximate cause. *See* Restatement (Third) of Torts: Liab. for Physical & Emotional Harm § 29 (Am. L. Inst. 2010) (limiting liability to harms within the scope of risks that made the actor's conduct tortious); *id.* § 34 (providing that an intervening act is not superseding when it is within the scope of the risks that made the actor's conduct tortious).

[123] *See Grokster*, 545 U.S. at 936-37; *Napster*, 239 F.3d at 1020-21.

[124] Because several of these case studies involve ongoing litigation, the analysis relies on current procedural rulings and factual allegations to illustrate the framework's utility, rather than predicting ultimate judicial outcomes.

[125] *See infra* Part IV.A (autonomous vehicle tort liability); Part IV.B (algorithmic housing discrimination); Part IV.C (algorithmic hiring discrimination); Part IV.D (AI-assisted policing and the Fourth Amendment); Part IV.E (algorithmic pricing coordination). For the common

- **Autonomous Vehicle Collision (Tort).** *Puzzle:* The "empty driver's seat" obscures fault in an accident. *Doctrinal Hook:* Risk-utility design defect.[126] *OA/OAG Contribution:* Shifts the focus from the absent driver to the developer's culpable design choices, treating the AI's defective post-collision logic as the basis for a product defect claim.

- **Algorithmic Housing Discrimination (FHA).** *Puzzle:* A "facially neutral algorithm" hides discriminatory impact. *Doctrinal Hook:* Disparate-impact liability.[127] *OA/OAG Contribution:* Traces discriminatory impact to the vendor's model architecture and frames the landlord's decision to deploy the tool as a distinct, negligent act, clarifying the basis for shared liability.

- **Biased Hiring Algorithm (Title VII/ADEA).** *Puzzle:* A vendor claims immunity as a "neutral toolmaker." *Doctrinal Hook:* Liability as an agent of the employer.[128] *OA/OAG Contribution:* Pierces the neutral-vendor defense by analyzing the AI's objective function, reframing the vendor as an active participant in the discriminatory decision-making process.

- **Algorithmic Policing (Constitutional Law).** *Puzzle:* Public-private entanglement complicates constitutional claims. *Doctrinal Hook: Monell* municipal liability. *OA/OAG Contribution:* Maps the distinct but linked failures of the private vendor (for providing an unreliable system) and the city (for its policy of uncritical reliance), making the resulting systemic failure legally legible.

- **AI-Facilitated Price-Fixing (Antitrust).** *Puzzle:* An "algorithmic cartel" obscures the necessary "agreement" to collude. *Doctrinal Hook:* Sherman Act §1 conspiracy. *OA/OAG Contribution:* Locates the "agreement" in the act of ceding pricing authority to a

---

structural challenge these cases present—the AI's autonomous decision-making interposing sufficient causal and epistemic distance to defeat existing doctrines—*see supra* notes 22–24 and accompanying text.

[126] *See* RESTATEMENT (THIRD) OF TORTS: PRODS. LIAB. § 2(b) (Am. L. Inst. 1998) (providing that a product is defective in design when foreseeable risks could have been reduced by a reasonable alternative design).

[127] *See* Tex. Dep't of Hous. & Cmty. Affs. v. Inclusive Cmtys. Project, Inc., 576 U.S. 519, 539–40 (2015) (recognizing disparate-impact claims under the Fair Housing Act).

[128] *See* 42 U.S.C. § 2000e(b) (2018) (defining "employer" under Title VII to include "any agent" of the employer); 29 U.S.C. § 630(b) (2018) (same, under the ADEA).

centralized AI whose very design and objective function are engineered to enforce the conspiracy.

### A. Autonomous Vehicle Collision (Tort & Products Liability)

The October 2023 Cruise robotaxi incident presents a classic products liability and negligence case, but one that highlights the "empty driver's seat" puzzle: in the absence of a human driver to blame, traditional negligence analysis can be obscured, creating a potential liability vacuum.[129] OAG is designed to fill this vacuum by shifting the focus from the non-existent driver to the legally responsible designer.[130]

In the incident, after a human driver in another vehicle struck a pedestrian, the victim was thrown into the path of the Cruise Autonomous Vehicle (AV).[131] The autonomous system failed to correctly identify the nature of the collision, misclassifying it as a lateral collision rather than one involving a vulnerable road user directly in its path.[132] Instead of remaining stationary, it initiated a pre-programmed "pullover" maneuver, dragging the trapped victim approximately 20 feet (and reaching approximately 7 mph before the final stop) and causing severe injuries.[133]

---

[129] For the core facts of the Cruise incident, *see supra* note 40 and accompanying text; for detailed analysis, *see infra* notes 134–44 and accompanying text. The absence of a human driver eliminates the traditional focal point for negligence analysis. *See* Vladeck, *supra* note 42, at 141–45 (analyzing the liability gap that results when autonomous systems operate without identifiable human principals).

[130] *See Fluid Agency*, *supra* note 7, Part VI (proposing the analysis of an AI system's operational characteristics as a method for tracing accountability to human design choices); RESTATEMENT (THIRD) OF TORTS: PRODS. LIAB. § 2(b) (Am. L. Inst. 1998) (providing the doctrinal basis for this approach by assessing product defectiveness through the designer's choices rather than end-user conduct).

[131] *See* Quinn Emanuel Report, *supra* note 40, at 1 (describing the sequence of events in which a human-driven vehicle struck the pedestrian, who was then thrown into the path of the Cruise AV).

[132] *See* NHTSA Safety Recall Report, *supra* note 40, at 2 (stating that the Cruise ADS "inaccurately characterized the collision as a lateral collision and commanded the AV to attempt to pull over out of traffic . . . rather than remaining stationary").

[133] *See* CAL. DEP'T OF MOTOR VEHICLES, ORDER OF SUSPENSION TO CRUISE LLC, *supra* note 40 (describing that the AV "traveled approximately 20 feet and reached a speed of 7 mph" while the pedestrian was underneath the vehicle).

NHTSA's recall filing and Cruise's board-commissioned independent counsel report confirmed that the AV's software misclassified the event, triggering the maneuver.[134]

OAG helps organize the relevant evidence. The key nodes are "Cruise LLC" (Juridical Person) and its "Cruise ADS" (AI Agent).[135] The decisive edge is the one from Cruise LLC to its AI.[136] The weight on this edge is justified by a rigorous analysis of OA's pillars.

First, the system's *Safety Architecture (Pillar 3)* was fundamentally defective, a fact later confirmed in a federal safety recall filed by Cruise itself.[137] The National Highway Traffic Safety Administration (NHTSA) recall report acknowledged that the software inaccurately characterized the collision as a "lateral collision," causing it to attempt a pull-over maneuver instead of remaining stationary and thereby creating an unreasonable risk of injury.[138] This "brittle" design stands in stark contrast to feasible industry alternatives.[139] For example, both

---

[134] For the DMV suspension and core facts, *see* CAL. DEP'T OF MOTOR VEHICLES, ORDER OF SUSPENSION TO CRUISE LLC, *supra* note 40 (describing that, post-stop, "the AV attempted to perform a pullover maneuver while the pedestrian was underneath," that it "traveled approximately 20 feet and reached a speed of 7 mph," and concluding that continued testing "poses an unreasonable risk to the public"); *see also* CAL. DEP'T OF MOTOR VEHICLES, DMV STATEMENT ON CRUISE LLC SUSPENSION (Oct. 24, 2023), https://www.dmv.ca.gov/portal/news-and-media/dmv-statement-on-cruise-llc-suspension/ [https://perma.cc/TF24-UAG8]; CAL. PUB. UTILS. COMM'N, CLASS P CHARTER-PARTY PERMIT NO. TCP0039080-P FOR CRUISE LLC 1 (rev. May 30, 2024) (showing "DEPLOYMENT PROGRAM (SUSPENDED EFFECTIVE OCTOBER 24, 2023)"), https://www.cpuc.ca.gov/-/media/cpuc-website/divisions/consumer-protection-and-enforcement-division/documents/tlab/av-programs/cruise-tcp-p-consolidated-rev20240530.pdf [https://perma.cc/Q34M-JU8J].

[135] For products-liability purposes, Cruise is the relevant defendant as the designer and manufacturer of the ADS software stack that controlled the vehicle's post-collision behavior.

[136] The OAG focuses on this edge because the AI agent node is not a locus of liability but a conduit through which the developer's design choices manifest as harm. *See supra* note 109 and accompanying text (establishing that an AI node's operational profile informs culpability analysis of upstream human and juridical person nodes).

[137] *See* NHTSA Safety Recall Report, *supra* note 40, at 1–2 (documenting a voluntary safety recall of 950 Cruise autonomous vehicles to correct the defective post-collision response software); *supra* notes 76–84 and accompanying text (developing safety architecture as a functional proxy for the standard of care).

[138] NHTSA Safety Recall Report, *supra* note 40, at 2 '("The Cruise ADS inaccurately characterized the collision as a lateral collision and commanded the AV to attempt to pull over out of traffic … rather than remaining stationary.").

[139] *See* RESTATEMENT (THIRD) OF TORTS: PRODS. LIAB. § 2(b) (Am. L. Inst. 1998) (providing that a product is defective in design when foreseeable risks could have been reduced by a reasonable

Waymo and Mobileye have publicly documented safety cases specifying that their vehicles enter a *minimal risk condition* (i.e., a safe stop) after a collision, rather than attempting to continue driving—a far more prudent design.[140] The existence of these reasonable alternative designs supports a strong inference of breach—evidence that a safer post-collision protocol was feasible and that Cruise's failure to implement one constitutes a design defect.

Second, the harm was a foreseeable consequence under *Predictive Processing (Pillar 2)*. The AI's own sensors detected the initial impact, giving it knowledge that a severe collision had occurred.[141] The risk of a pedestrian being under or near a vehicle after a severe collision is not merely obvious; it is a critical edge case so foreseeable that it became the subject of a federal recall.[142] Moreover, the developer's constructive knowledge of this failure mode is undeniable, as its own post-incident investigation and subsequent federal recall filing confirmed the software's inability to correctly handle the pedestrian-under-vehicle scenario.[143] The developer's failure to program for this high-risk hazard was a clear breach of its duty.

---

alternative design); *cf. The T.J. Hooper*, 60 F.2d at 740 (holding that an entire industry's failure to adopt available safety technology does not excuse the defendant).

[140] WAYMO LLC, EMERGENCY RESPONSE GUIDE AND LAW ENFORCEMENT INTERACTION PROTOCOL 12 (Oct. 2023) ("In that event, the vehicle will brake until it reaches a full stop."), https://storage.googleapis.com/waymo-uploads/files/documents/first-responders/Waymo%20Emergency%20Response%20Guide%20and%20Law%20Enforcement%20Interaction%20Protocol%20%28October%202023%29%20-%20240122.pdf [https://perma.cc/B36S-GLN6]; MOBILEYE, A SAFETY ARCHITECTURE FOR SELF-DRIVING SYSTEMS 10 (2024) ("Each … potential failure comes with a Minimal Risk Maneuver (MRM)."), https://static.mobileye.com/website/us/corporate/files/SDS_Safety_Architecture.pdf [https://perma.cc/4WN9-SLWM].

[141] *See* Quinn Emanuel Report, *supra* note 40, at 12–15 (confirming that the vehicle's onboard sensors detected and tracked the pedestrian throughout the collision sequence). For the principle that such machine-generated data imputes constructive knowledge to the developer, *see supra* notes 66–75 and accompanying text (developing predictive processing as a functional proxy for foreseeability); RESTATEMENT (THIRD) OF TORTS: PRODS. LIAB. § 2 cmt. m (Am. L. Inst. 1998) (charging sellers with knowledge of what reasonable testing would reveal).

[142] *See* NHTSA Safety Recall Report, *supra* note 40, at 2 (identifying the post-collision pullover maneuver as the specific safety defect necessitating a recall of 950 vehicles). *Cf.* RESTATEMENT (THIRD) OF TORTS: PRODS. LIAB. § 2 cmt. m (Am. L. Inst. 1998) (charging sellers with knowledge of what reasonable testing would reveal).

[143] The Quinn Emanuel report confirms the vehicle's sensors detected and tracked the pedestrian leading up to the initial impact. It also notes the pedestrian was partially occluded from some sensors immediately prior to impact. The decisive failure, however, was not in perception but in

The OAG thus supports design-defect and corporate-negligence theories under the risk-utility framework, re-centering the analysis on the developer's concrete design choices rather than the AV's inscrutable post-collision 'choice.' By doing so, it provides a clear, evidence-based solution to the empty driver's seat puzzle.[144]

**B. Algorithmic Housing Discrimination (Civil Rights & Disparate Impact)**

The use of automated decision-making systems in housing presents a significant challenge for civil rights enforcement, centered on the "facially neutral algorithm" puzzle.[145] Under the Fair Housing Act (FHA), a housing practice can be unlawful if it has an unjustified discriminatory effect, regardless of intent.[146] The evidentiary challenge is to demonstrate how a seemingly objective algorithm, which uses neutral inputs like credit history, produces a discriminatory disparate impact.[147] The settled case of *Louis v. SafeRent Solutions, LLC*

---

the software's post-collision classification of the event and its subsequent decision-making logic. *See* QUINN EMANUEL REPORT, *supra* note 40, at 1, 12-15.

[144] For subsequent enforcement and compliance actions, *see, e.g.,* CAL. PUB. UTILS. COMM'N, Decision 24-07-004 (July 25, 2024) (approving a $112,500 penalty for incomplete disclosures); NAT'L HIGHWAY TRAFFIC SAFETY ADMIN., CONSENT ORDER (Sep. 30, 2024), https://www.nhtsa.gov/press-releases/consent-order-cruise-crash-reporting [https://perma.cc/PHM7-B4Y4] ($1.5 million for failing to fully report the dragging event); Press Release, U.S. Att'y's Office, N.D. Cal., *Cruise Admits to Submitting a False Report to Influence a Federal Investigation and Agrees to Pay $500,000* (Nov. 14, 2024), https://www.justice.gov/usao-ndca/pr/cruise-admits-submitting-false-report-influence-federal-investigation-and-agrees-pay [https://perma.cc/ZT7C-5M5W] ($500,000 criminal penalty and compliance program). GM later announced it would cease funding Cruise's robotaxi development. *See, e.g., GM to Exit Cruise Robotaxi Business*, REUTERS (Dec. 10, 2024), https://www.reuters.com/business/autos-transportation/general-motors-drop-development-cruise-robotaxi-2024-12-10/ [https://perma.cc/M3PB-6YDW].

[145] *See, e.g.*, Solon Barocas & Andrew D. Selbst, *Big Data's Disparate Impact*, 104 Cal. L. Rev. 671, 677–93 (2016) (analyzing how facially neutral algorithmic processes produce discriminatory outcomes through biased training data and proxy variables, even absent discriminatory intent).

[146] *See Inclusive Cmtys. Project*', 576 U.S. at 539-40 (recognizing disparate-impact liability under the FHA).

[147] *See* Barocas & Selbst, *supra* note 145, at 688–92 (explaining how facially neutral algorithmic inputs—including credit history and geographic indicators—systematically produce discriminatory outcomes by serving as proxies for protected class membership through "redundant encodings," even absent discriminatory intent).

illustrates this problem and provides a springboard for demonstrating the utility of OA and OAG.[148]

The plaintiffs, a class of rental applicants, alleged that SafeRent's automated tenant-screening service disproportionately denied housing to Black and Hispanic applicants, particularly those using federal Section 8 housing vouchers.[149] The lawsuit claimed that SafeRent's AI-generated "SafeRent Score" was discriminatory because its design overemphasized credit history and failed to properly account for their guaranteed government income.[150] The complaint alleged that this design caused a stark disparate impact, making applicants with vouchers disproportionately likely to be denied housing.[151] Voucher status is not itself a protected characteristic under the FHA; rather, plaintiffs alleged that the model's reliance on credit metrics produced a disparate impact based on race and national origin.[152] The U.S. Department of Justice underscored the significance of the legal theory by filing a Statement of Interest in the case, affirming that the FHA applies to algorithmic screening systems.[153]

The OAG analysis begins with the edge connecting "SafeRent Solutions" (Juridical Person) to its "Screening AI", typed as "Tenant Scoring Model Design."[154] The weight of this edge can be assessed using the OA pillars.[155] The system's *Safety Architecture (Pillar 3)* is the central issue in a disparate impact claim.[156] The plaintiffs alleged the AI's design was fundamentally flawed because it treated applicants with vouchers, who have a guaranteed government-backed income source, as higher risk if they had a poor credit history.[157] The feasibility of a reasonable alternative design is not merely theoretical; it is evidenced by the core terms of the eventual settlement.[158] Under those terms, for five years, SafeRent must withhold its "SafeRent Score" and any "accept/decline" recommendation for voucher applicants in its "affordable" housing model and, absent specific certification, in its "market" and "no-credit" models as well.[159] The prior failure to adopt such an alternative weighs heavily in the burden-shifting analysis.[160] Under *Predictive Processing (Pillar 2)*, the risk of disparate impact from such models was a foreseeable issue establishing SafeRent's constructive knowledge.[161]

---

[148] *See* Amended Complaint for Violations of the Fair Housing Act, Louis v. SafeRent Sols., LLC, No. 1:22-cv-10800-AK (D. Mass. Aug. 26, 2022), ECF No. 15.

[149] *See* Louis v. SafeRent Sols., LLC, 685 F. Supp. 3d 19, 26–28 (D. Mass. 2023) (summarizing the plaintiffs' class allegations that SafeRent's algorithmic tenant-screening service disproportionately denied housing to Black and Hispanic voucher holders).

[150] *See* Amended Complaint, *supra* note 148, at ¶¶ 30-31, 45-50.

[151] *Id.* at ¶¶ 4-5.

[152] Source of income is not a protected characteristic under the federal Fair Housing Act. *See* 42 U.S.C. § 3604(a) (2018) (prohibiting discrimination in housing on the basis of "race, color, religion, sex, familial status, or national origin"). For the plaintiffs' theory that the algorithm's facially neutral reliance on credit metrics functioned as a proxy for race and national origin, *see Louis*, 685 F. Supp. 3d at 39–41 (denying SafeRent's motion to dismiss and holding that the plaintiffs adequately alleged that credit-based screening criteria produced a racially disparate impact actionable under the FHA).

[153] *See* Statement of Interest of the United States at 1-2, Louis v. SafeRent Sols., LLC, No. 1:22-cv-10800-AK (D. Mass. Jan. 9, 2023), ECF No. 37.

[154] *See supra* notes 106–11 and accompanying text (defining OAG node types, edge attributes, and the open-ended edge typology, including "Design Choice" as an illustrative edge type).

[155] *See supra* notes 112–16 and accompanying text (explaining how the three OA pillars structure the qualitative assessment of each edge's weight).

Crucially, while the class settlement focused on the vendor, the case itself and broader legal principles confirm that liability extends to the housing providers who use such tools.[162]

---

[156] *See supra* notes 76–84 and accompanying text (developing Pillar 3). In the disparate-impact context, Pillar 3's inquiry into whether a feasible, less-risky design existed corresponds to the final step of the FHA's burden-shifting framework, under which a plaintiff may prevail by demonstrating a less discriminatory alternative that serves the defendant's legitimate interests. *See Inclusive Cmtys. Project*, 576 U.S. at 541–43; 24 C.F.R. § 100.500(c)(3) (2023).

[157] *See* Amended Complaint, *supra* note 148, ¶¶ 45–50 (alleging that the algorithm overweighted credit history and non-tenancy debts while failing to account for guaranteed government income from housing vouchers).

[158] *See* RESTATEMENT (THIRD) OF TORTS: PRODS. LIAB. § 2(b) (Am. L. Inst. 1998) (providing that a design is defective when foreseeable risks could have been reduced by a reasonable alternative design); *supra* notes 76–84 and accompanying text (developing Pillar 3). That SafeRent itself implemented such an alternative in settlement, *see infra* note 159, is compelling evidence of feasibility. *Cf. The T.J. Hooper*, 60 F.2d at 740 (holding that the availability of a known safety measure creates a duty to adopt it regardless of industry custom).

[159] *See* Class Action Settlement Agreement & Release §§ 3.5.1-3.5.3, Louis v. SafeRent Sols., LLC, No. 1:22-cv-10800-AK (D. Mass. Mar. 28, 2024), ECF No. 114-1; Order Granting Final Approval of Class Action Settlement at 2, *id.* (Nov. 20, 2024), ECF No. 135 (retaining jurisdiction for five years to enforce the practice changes).

[160] *See* 24 C.F.R. § 100.500(c)(3) (2023) (providing that a challenged practice with a discriminatory effect violates the FHA if a less discriminatory alternative serves the same legitimate objective); *Inclusive Cmtys. Project*, 576 U.S. at 541–43.

[161] *See* Statement of Interest, *supra* note 153, at 6-10.

[162] *See* Class Action Settlement Agreement & Release, *supra* note 159, §§ 1.17–1.18 (defining the settlement classes as voucher holders denied housing based on their SafeRent Score), 3.6.1 (releasing SafeRent from all claims); *Louis*, 685 F. Supp. 3d at 34–36 (holding that SafeRent is subject to the FHA and that a housing provider cannot "use an intermediary to take

This is where the OA framework provides a critical evidentiary bridge. Under the FHA, housing providers remain directly and vicariously liable for discriminatory housing decisions, including those made through agents or third-party tools.[163] The court in *Louis* affirmed this principle by denying the landlord-defendant's motion to dismiss the FHA claims.[164]

The OAG makes this shared responsibility legible. It would add a "Landlord" node with a causal edge to the "Screening AI," typed as "Selection and Use of Screening Tool." The weight of this edge is determined by the landlord's own breach of care. By choosing a tool with a demonstrably flawed OA profile—a *Safety Architecture (Pillar 3)* lacking reasonable safeguards, despite the *foreseeable* disparate impact (*Pillar 2*)—the landlord commits its own negligent act. OA thus provides the evidentiary basis to hold landlords accountable for their selection and use of discriminatory tools, re-anchoring the non-delegable duty to prevent discrimination squarely on the party that controls access to the property.

**C. Biased Hiring Algorithm (Civil Rights & Corporate Agency)**

The ongoing lawsuit in *Mobley v. Workday, Inc.* highlights the "neutral vendor" puzzle: the challenge of apportioning liability to a developer who claims to be a mere passive toolmaker.[165] OAG addresses this defense by analyzing the functional design of the tool itself.

Specifically, the software-as-a-service company Workday faces a class action alleging its AI-powered hiring tools have a disparate impact on applicants based on race, age, and

---

discriminatory and prohibited actions on its behalf and defeat the purpose of the FHA" (quoting Conn. Fair Hous. Ctr. v. CoreLogic Rental Prop. Sols., LLC, 369 F. Supp. 3d 362, 374–75 (D. Conn. 2019)).

[163] *See* 24 C.F.R. § 100.7(a)-(b) (2023) (establishing direct and vicarious liability for discriminatory housing practices); *see also* U.S. DEP'T OF HOUS. & URB. DEV., GUIDANCE ON APPLICATION OF THE FAIR HOUSING ACT TO THE SCREENING OF APPLICANTS FOR RENTAL HOUSING 3-4 (Apr. 29, 2024), https://www.fairhousingnc.org/wp-content/uploads/2024/08/FHEO_Guidance_on_Screening_of_Applicants_for_Rental_Housing.pdf [https://perma.cc/KZ25-7XDS] (advising that housing providers remain liable under the FHA when using tenant-screening tools that produce discriminatory outcomes).

[164] *See Louis*, 685 F. Supp. 3d at 31–34 (denying Metropolitan's motion to dismiss and holding that Louis established standing to bring FHA claims against the housing provider that relied on SafeRent's screening tool).'

[165] *See* Order Granting in Part and Denying in Part Motion to Dismiss at 5, 8, Mobley v. Workday, Inc., No. 23-cv-00770-RFL (N.D. Cal. July 12, 2024), ECF No. 80 [hereinafter *Mobley* MTD Order] (describing Workday's contention "that, as a software vendor, it is not a covered entity" under the federal anti-discrimination statutes, and observing that accepting this argument "would allow companies to escape liability for hiring decisions by saying that function has been handed over to someone else (or here, artificial intelligence)").

disability.[166] In a key July 2024 order, the court issued a split ruling.[167] It rejected the theory that Workday was an "employment agency" but allowed disparate impact claims to proceed on the theory that Workday acted as an "agent" of employers, finding it plausible that employers delegated traditional hiring functions to Workday's tools, which were "participating in the decision-making process by recommending some candidates to move forward and rejecting others."[168] A subsequent May 2025 order preliminarily certified an age discrimination collective based on a "unified policy: the use of Workday's AI recommendation system to score, sort, rank, or screen applicants," identifying specific tools like the Candidate Skills Match (CSM) and Workday Assessment Connector (WAC).[169] A later order further expanded the set of implicated tools to include additional AI features like Spotlight and Fetch.[170]

An OAG analysis clarifies Workday's role. The graph's critical edge is the one from "Workday Inc." (Juridical Person) to its "Workday Screen AI" (AI Agent).

First, the risk of discriminatory outcomes was eminently foreseeable under *Predictive Processing (Pillar 2)*. A sophisticated vendor like Workday had far more than general knowledge of this risk.[171] Federal regulators had issued guidance warning that *employers* remain liable for discrimination caused by the algorithmic tools they use, putting vendors on notice of discrimination risks and articulating expectations relevant to *Safety Architecture (Pillar 3)*.[172] Moreover, the EEOC underscored the risk to vendors themselves by filing an amicus brief in this very case, arguing that an AI vendor can be held liable as an agent under federal anti-discrimination law.[173] This direct regulatory intervention put Workday on notice of the foreseeable risks.[174]

---

[166] *Id.* at 1–3 (describing Mobley's claims under Title VII, the ADEA, and the ADA alleging that Workday's algorithm-based applicant screening tools discriminated on the basis of race, age, and disability).

[167] *Id.* at 1, 20 (granting the motion in part and denying it in part).

[168] *Id.* at 9–10 (finding it plausible that Workday's customers delegated traditional hiring functions to Workday's algorithmic tools, which were "participating in the decision-making process by recommending some candidates to move forward and rejecting others"), 12–13 (dismissing the employment agency theory).

[169] *See* Order Granting Preliminary Collective Certification at 2-4, 10-12, *Mobley*, No. 23-cv-00770-RFL (N.D. Cal. May 16, 2025), ECF No. 128 (conditionally certifying an ADEA collective and describing a "unified policy" using Workday's AI screening tools).

[170] *See* Order re Opt-In Plan at 1, 3, *Mobley*, No. 23-cv-00770-RFL (N.D. Cal. Aug. 28, 2025), ECF No. 166 (directing production of a list of employers that used Workday "AI features (including CSM, Spotlight and Fetch)").

[171] *See supra* notes 66–75 and accompanying text (developing Pillar 2's constructive-knowledge framework). Workday's own tools allegedly embed "artificial intelligence ('AI') and machine learning ('ML') into its algorithmic decision-making tools, enabling these applications to make hiring decisions." *Mobley* MTD Order, *supra* note 165, at 2; *see also* Barocas & Selbst, *supra* note 145, at 677–93 (documenting the well-established risk that facially neutral algorithmic processes produce discriminatory outcomes through biased training data and proxy variables).

Second, under *Goal-Directedness (Pillar 1)*, the AI's core objective—to score, sort, and rank candidates based on their similarity to historically "successful" profiles—becomes a discriminatory proxy when the underlying data is biased. This is not a theoretical flaw; peer-reviewed research on commercial hiring algorithms has empirically demonstrated that automated screening systems often reproduce historical biases, particularly in models that, like Workday's, score candidates based on their similarity to historically successful profiles.[175]

The AI's engineered goal, when combined with the foreseeable nature of biased training data, illustrates a system primed for discriminatory impact. This OAG analysis gives courts a structured method to trace this inherent risk to a vendor's design choices.[176] The analysis, however, does not absolve other actors. Just as landlords retain a non-delegable duty in housing, employers who choose to adopt such tools retain their own duty to ensure their hiring practices are non-discriminatory. A parallel OAG analysis of the employer's decision to adopt the tool would weigh the reasonableness of their due diligence in light of the foreseeable risks documented in public regulatory guidance, creating a clear basis for apportioning liability. The framework thus resolves the "neutral vendor" puzzle by providing a concrete evidentiary method to distinguish passive toolmaking from the active participation in screening that grounds liability for vendors and users alike.[177]

---

[172] *See* U.S. EQUAL EMP. OPPORTUNITY COMM'N, SELECT ISSUES: ASSESSING ADVERSE IMPACT IN SOFTWARE, ALGORITHMS, AND ARTIFICIAL INTELLIGENCE USED IN EMPLOYMENT SELECTION PROCEDURES UNDER TITLE VII (May 18, 2023), https://www.eeoc.gov/select-issues-assessing-adverse-impact-software-algorithms-and-artificial-intelligence-used [https://perma.cc/WJ3K-W7RD]; U.S. DEP'T OF JUSTICE, C.R. DIV., ALGORITHMS, ARTIFICIAL INTELLIGENCE, AND DISABILITY DISCRIMINATION IN HIRING (May 12, 2022), https://www.ada.gov/resources/ai-guidance/ [https://perma.cc/5DCZ-V7PW].

[173] Brief for the U.S. Equal Employment Opportunity Commission as Amicus Curiae Supporting Plaintiffs at 6-13, *Mobley*, No. 23-cv-00770-RFL (N.D. Cal. Apr. 9, 2024), ECF No. 60-1.

[174] *See supra* notes 66–75 and accompanying text (Pillar 2); *supra* notes 76–84 and accompanying text (Pillar 3). The EEOC and DOJ guidance cited in the two preceding notes, together with the EEOC's direct intervention as amicus in this case, collectively establish both that the risk of algorithmic discrimination was foreseeable and that an evolving standard of care for AI hiring vendors was being articulated by federal regulators.

[175] *See* Manish Raghavan, Solon Barocas, Jon Kleinberg & Karen Levy, *Mitigating Bias in Algorithmic Hiring: Evaluating Claims and Practices*, *in* PROC. 2020 CONF. ON FAIRNESS, ACCOUNTABILITY, AND TRANSPARENCY 469, 471 (2020) (finding that automated screening often serves as a "veneer of objectivity" for existing, biased hiring practices); *see also* Barocas & Selbst, *supra* note 145, at 677–93 (documenting how algorithmic processes produce discriminatory outcomes through biased training data and proxy variables).

[176] *See supra* notes 106–16 and accompanying text (defining OAG node types and edge attributes, and explaining how the three OA pillars structure the qualitative assessment of each edge's causal weight).

**D. Algorithmic Policing Alert (Constitutional Law & State Action)**

The case of *Williams v. City of Chicago* exemplifies the "public-private entanglement" puzzle: untangling a complex web of actors to establish liability for a constitutional violation.[178] The OAG is well-suited to map this multi-actor system and make its systemic failures legally legible.

In *Williams*, a man spent nearly a year in jail after being arrested based on a ShotSpotter alert that was algorithmically generated and then manually reviewed by a company analyst.[179] After prosecutors dismissed the criminal case for insufficient evidence to meet their burden of proof, Williams filed a federal civil rights lawsuit against the City of Chicago.[180] In a key ruling, the court denied the City's motion to dismiss in substantial part, finding it plausible that an arrest based on an unreliable alert could violate the Fourth Amendment and allowing the *Monell* claims to proceed.[181]

---

[177] *Cf.* City of L.A., Dep't of Water & Power v. Manhart, 435 U.S. 702, 718 n.33 (1978) ("[A]n employer can[not] avoid his responsibilities by delegating discriminatory programs to corporate shells."); *Mobley* MTD Order, *supra* note 165, at 10–11 (warning that drawing "an artificial distinction between software decisionmakers and human decisionmakers would potentially gut anti-discrimination laws in the modern era"). For the parallel principle in housing, *see Louis*, 685 F. Supp. 3d at 41.

[178] *See* 42 U.S.C. § 1983 (2018). Municipal liability under § 1983 requires proof that a deprivation of constitutional rights was caused by a government "policy or custom." *See* Monell v. Dep't of Soc. Servs., 436 U.S. 658, 694 (1978).

[179] *See* Garance Burke, Martha Mendoza, Juliet Linderman & Michael Tarm, *How AI-Powered Tech Landed Man in Jail with Scant Evidence*, AP News (Aug. 19, 2021), https://apnews.com/article/artificial-intelligence-algorithm-technology-police-crime-7e3345485aa668c97606d4b54f9b6220 [https://perma.cc/25KS-UPGM] (reporting that a man was jailed based on a ShotSpotter alert before prosecutors dismissed the case for insufficient evidence).

[180] *See* Complaint at ¶¶ 2, 5, 8, Williams v. City of Chicago, No. 1:22-cv-03773 (N.D. Ill. filed July 21, 2022).

The OAG analysis highlights several interactions that, collectively, suggest a systemic problem with constitutional implications.[182] First, the edge from "SoundThinking Inc." (the private vendor).[183] The system's unreliability, a key component of *Predictive Processing (Pillar 2)*, was not merely alleged but documented by the City's own neutral oversight body.[184] A 2021 report from the Chicago Office of the Inspector General (OIG) found that only 9.1% of police responses to ShotSpotter alerts resulted in documented evidence of a gun-related offense.[185]

Furthermore, the vendor's *Goal-Directedness (Pillar 1)* is implicated by a defamation ruling in which the court discussed record testimony.[186] A Delaware court, reviewing the record at the motion-to-dismiss stage, noted that "it is apparent… that there is a pattern of alterations, and that these alterations sometimes come by request of police departments," suggesting a corporate purpose that prioritizes client satisfaction.[187] These factors establish strong evidence supporting constructive knowledge for both the vendor and the city.[188]

---

[181] *See* Order at 12, 15-17, *Williams*, No. 1:22-cv-03773 (N.D. Ill. Sep. 29, 2023), ECF No. 183 (denying the City's motion to dismiss *Monell* claims and finding it plausible that Williams's arrest based on a ShotSpotter alert violated the Fourth Amendment).

[182] *See supra* notes 106–16 and accompanying text (defining OAG node types and edge attributes).

[183] SoundThinking, Inc. (formerly ShotSpotter, Inc.) is the developer and operator of the acoustic gunshot-detection system at issue. *See* Burke *et al.*, *supra* note 179.

[184] *See supra* notes 66–75 and accompanying text (developing Pillar 2's constructive-knowledge framework).

[185] *See* CITY OF CHI. OFF. OF INSPECTOR GEN., CHICAGO POLICE DEPARTMENT'S USE OF SHOTSPOTTER TECHNOLOGY 3 (Aug. 24, 2021), https://igchicago.org/wp-content/uploads/2021/08/Chicago-Police-Departments-Use-of-ShotSpotter-Technology.pdf. The vendor disputes the OIG's methodology and findings. *See* SoundThinking, *Why the Chicago OIG Report is Misunderstood*, SOUNDTHINKING (May 20, 2024), https://www.soundthinking.com/blog/soundthinkings-commentary-on-chicago-oig-report/ [https://perma.cc/9H8P-WZ6G].

[186] *See supra* notes 63–65 and accompanying text (developing Pillar 1's goal-directedness analysis).

Second, and critically, the edge from the "City of Chicago" to its "Chicago Police Department (CPD) Officers" reflects a flawed *Safety Architecture (Pillar 3)* at the systemic level. The City's policies and practices encouraged officers to treat ShotSpotter alerts as sufficient grounds for police action, often without meaningful independent corroboration, a practice in tension with existing legal guidance. The governing appellate court had already questioned whether a single ShotSpotter alert, by itself, could amount to reasonable suspicion.[189] In Williams's criminal case, prosecutors sought dismissal after concluding the evidence was "insufficient to meet our burden of proof."[190] This policy of uncritical reliance, in the face of the OIG's documented low evidentiary-yield rate, could support a finding of deliberate indifference to the obvious risk of Fourth Amendment violations.

Later developments underscore the salience of these concerns. In February 2024, the City of Chicago announced it would terminate its contract with ShotSpotter, with use scheduled to end by September 22, 2024 (followed by a short wind-down period).[191] More decisively, in August 2025, the city settled the related class action lawsuit, agreeing as a matter of policy that a ShotSpotter alert, standing alone, does not provide justification for a stop or pat-down.[192] Williams's individual case was severed and remains pending.[193]

---

[187] *See* ShotSpotter, Inc. v. VICE Media, LLC, No. N21C-10-082 SKR, at 17-18 (Del. Super. Ct. June 30, 2022) (granting VICE's motion to dismiss and noting record testimony indicating a "pattern of alterations," sometimes "per the instruction of the customer"). The court quoted testimony from ShotSpotter analyst Paul Greene in two separate criminal cases where he confirmed that classifications are changed "per the instruction of the customer" and that police requests to search for additional sounds happen on a "semiregular basis." *Id.*

[188] *See supra* notes 66–75 and accompanying text (developing Pillar 2's constructive-knowledge framework). The OIG findings cited *supra* note 185, together with the record testimony described *supra* note 187, constitute precisely the kind of discoverable evidence that Pillar 2 identifies as probative of constructive knowledge.

[189] *See* United States v. Rickmon, 952 F.3d 876, 881 (7th Cir. 2020) ("[W]e question whether a single ShotSpotter alert would amount to reasonable suspicion.").

[190] *See* Burke et al., *supra* note 180.

[191] *See* City of Chicago, Statement on ShotSpotter Contract (Feb. 13, 2024); *see also* Diba Mohtasham, *Chicago will drop controversial ShotSpotter gunfire detection system*, NPR (Feb. 15, 2024), https://www.npr.org/2024/02/15/1231394334/shotspotter-gunfire-detection-chicago-mayor-dropping [https://perma.cc/D2P5-K8HR] (reporting on Chicago's decision not to renew its ShotSpotter contract).

Viewed through the OAG, the constitutional harm was not a random failure but the foreseeable result of a chain of culpable human decisions. The framework makes this chain legible, allowing a court to move beyond the opaque algorithm and assign specific weight to the critical causal links: the vendor's design choices that produced an unreliable system, and the city's deliberate policy of uncritical reliance on it. The OAG thus provides a structured method to trace the path from flawed design and flawed policy to concrete constitutional injury.

**E. Algorithmic Cartel (Antitrust & Systemic Economic Harm)**

The challenge of AI-driven liability extends beyond physical and dignitary harms to encompass systemic economic injuries, as exemplified by the ongoing federal and private antitrust actions targeting RealPage, Inc. The case presents the "algorithmic cartel" puzzle: how to establish liability when an AI system, rather than human actors, appears to be the architect of a price-fixing scheme.

In a traditional price-fixing conspiracy under Section 1 of the Sherman Act, liability hinges on proving an agreement—a "meeting of the minds"—among competitors.[194] The evidentiary challenge in the RealPage litigation is to establish the existence of such an agreement when the alleged co-conspirators (landlords) cede their pricing authority to a centralized AI, which then acts as the hub in a "hub-and-spoke" conspiracy.[195]

---

[192] *See* Press Release, Roderick & Solange MacArthur Justice Ctr., *Chicago Agrees to Settle Lawsuit Challenging its Use of ShotSpotter* (Aug. 15, 2025) (announcing settlement terms in *Ortiz v. City of Chicago*, No. 1:22-cv-03773).

[193] *See id*. The *Ortiz* class action was consolidated with the *Williams* case before Williams's individual damages claim was severed to proceed separately as *Williams v. City of Chicago*, No. 1:24-cv-01396.

[194] *See* 15 U.S.C. § 1 (2018) ("Every contract, combination in the form of trust or otherwise, or conspiracy, in restraint of trade or commerce among the several States . . . is declared to be illegal."); Monsanto Co. v. Spray-Rite Serv. Corp., 465 U.S. 752, 764 (1984) (requiring evidence that "tends to exclude the possibility" of independent action to establish a § 1 conspiracy).

[195] *See* Complaint, United States v. RealPage, Inc., No. 1:24-cv-00710 (M.D.N.C. Aug. 23, 2024) ¶¶ 1-5 [hereinafter *RealPage* Complaint]; *see also In re* RealPage, Inc., Rental Software Antitrust Litig. (No. II), 709 F. Supp. 3d 478, 503 (M.D. Tenn. 2023) (denying motions to dismiss the multifamily claims while dismissing the student-housing complaint, analyzing the hub-and-spoke theory at the pleading stage, and describing plaintiffs' core allegation that competitors "delegate their rental price and supply decisions to a common decision maker, RealPage").

Specifically, the government alleges that RealPage orchestrated an unlawful price-fixing scheme through its revenue-management software, including YieldStar and AI Revenue Management (AIRM).[196] According to the complaint, competing landlords provided RealPage with nonpublic, competitively sensitive data—such as executed lease prices and forward-looking occupancy—which the algorithm aggregated to generate unit-level pricing recommendations that landlords adopted at supracompetitive levels.[197] Further, RealPage's design and workflows were engineered to reinforce adherence to its recommendations, for instance by requiring justifications for overrides and escalating disagreements to "pricing advisors."[198] In parallel, the government pleads a Section 2 theory, alleging RealPage's dominance in the commercial revenue-management-software market, reinforced by its data advantages.[199]

OAG maps influence and culpability in this multi-actor system. The graph's central edge connects "RealPage, Inc." (Juridical Person) to its "YieldStar/AIRM" software (AI Agent). This edge carries a *Heavy* weight, justified by a direct application of the OA pillars. First, the AI's *Goal-Directedness (Pillar 1)* is alleged to be inherently anti-competitive, as the complaint alleges that with this system, "RealPage replaces competition with coordination."[200] Second, this goal was achieved through *Predictive Processing (Pillar 2)* that relied on a flow of information—confidential rent rolls and occupancy data from competitors—that would be illegal for humans to share directly.[201] The complaint thus suggests that RealPage had constructive knowledge of the anti-competitive nature of this arrangement, as facilitating this data exchange was central to the AI's function.[202] Finally, the system's *Safety Architecture (Pillar 3)* allegedly functioned to reinforce adherence rather than prevent collusion, using features like "pricing advisors" to discourage deviation from its recommendations.[203]

---

[196] *See RealPage* Complaint, *supra* note 195, ¶¶ 3, 5, 14 (describing RealPage's software as replacing "competition with coordination" and identifying YieldStar, AIRM, and LRO as its three revenue management products).

[197] *See id.* ¶¶ 5-7, 15.

[198] *See id.* ¶¶ 61-63, 68-70.

[199] *See id.* ¶¶ 11, 14.

[200] *See id.* ¶ 3.

[201] *See id.* ¶ 15.

[202] *See supra* notes 66–75 and accompanying text (developing Pillar 2's constructive-knowledge framework); *RealPage* Complaint, *supra* note 195, ¶¶ 1–2, 6-7 (alleging that RealPage's stated purpose was to ensure landlords "move in unison versus against each other" and that it monitored compliance with its recommendations).

[203] *See supra* notes 76–84 and accompanying text (developing Pillar 3); *RealPage* Complaint, *supra* note 195, ¶¶ 68-70.

The RealPage litigation affirms this approach of anchoring the agreement in the AI's discoverable design. A federal court overseeing the private class-action litigation denied the defendants' motion to dismiss, holding that plaintiffs had plausibly alleged that ceding pricing authority to a centralized algorithm constituted an unlawful agreement.[204] Federal enforcement actions have targeted the same information flows, with the DOJ securing Proposed Final Judgments against major landlords—including Cortland Management, LLC and Greystar—that prohibit the use of revenue-management software relying on non-public competitor data.[205] Most recently, in November 2025, plaintiffs in the private litigation secured preliminary approval for settlements totaling $141.8 million with twenty-seven landlords, which included structural relief requiring them to cease sharing such data with RealPage.[206]

The framework thus addresses the "outsourced agreement" puzzle by re-anchoring the analysis in the discoverable design and function of the AI system itself. By deconstructing the AI's operational characteristics, OAG provides a method for tracing the alleged anti-competitive harm back to the concrete design choices of its developer. It allows a court to analyze the AI not as an inscrutable black box, but as a system whose engineered goal-directedness, reliance on sensitive data, and enforcement mechanisms provide the structured evidence needed to assess whether its adoption constituted an unlawful agreement under the Sherman Act.

## V. A Sword and a Shield: Re-Anchoring Liability to Human Actors

The case studies in the preceding Part demonstrate how OA can trace responsibility in specific scenarios. This Part now synthesizes those applications to show more broadly how OA functions as a dual-use tool—a *sword* that pierces legal immunities to hold irresponsible actors accountable and a *shield* that protects responsible innovation—to re-anchor liability in established legal doctrines.

The primary function of OA is apportioning fault among the multiple human actors in the causal chain. An AI's "OA profile"—its engineered *goal-directedness*, its embedded *predictive processing* capabilities, and its *safety architecture*—is a direct and discoverable manifestation of its developer's design choices and diligence. In our recurring hypothetical, an OA profile for CreatorBot that prioritizes rapid data acquisition over legal compliance serves as evidence of developer negligence for failing to build in adequate safeguards, or even recklessness for consciously disregarding the known risks of copyright infringement. This evidence provides the foundation for a products liability claim based on a design defect, shifting the legal inquiry from the AI's inscrutable "mind" to the developer's concrete and culpable design choices.

---

[204] *See In re RealPage, Inc.*, 709 F. Supp. 3d at 503, 510-12.

[205] *See* Proposed Final Judgment as to Defendant Cortland Management, LLC, United States v. RealPage, Inc., No. 1:24-cv-00710 (M.D.N.C. Jan. 7, 2025); Proposed Final Judgment as to Defendant Greystar Mgmt. Servs., LLC, *id.* (Aug. 8, 2025).

[206] *See* Order on Preliminary Approval of Settlements, *In re* RealPage, Inc., Rental Software Antitrust Litig. (No. II), No. 3:23-md-03071 (M.D. Tenn. Nov. 21, 2025) (preliminarily certifying a settlement class and approving 26 settlements with 27 defendants providing $141.8 million in monetary relief plus cooperation and injunctive terms); *see also* Joint Declaration of Interim Co-Lead & Liaison Counsel in Support of Preliminary Approval ¶ 5, *id.*, Doc. No. 1250 (summarizing monetary and non-monetary settlement terms).

Crucially, however, *OA is not a strict liability regime.*[207] If a harm results from a truly unforeseeable emergent behavior—one that a reasonable developer could not have predicted and that was not enabled by a flawed or brittle safety architecture—the framework correctly finds no fault.[208] This provides a vital safe harbor, protecting good-faith innovation from liability for the genuinely unpredictable and ensuring that accountability remains tethered to foreseeable risk.[209]

The framework also clarifies the user's role, allowing for a fair apportionment of liability and identifying when a user's intervening actions limit the scope of a developer's responsibility.[210] A user's culpability is measured *relative to the AI's known or knowable OA*

---

[207] While the severity of unmappability may point some toward strict liability regimes, OA provides a structured method to make fault-based inquiries tractable again, preserving traditional tort principles. *See Fluid Agency*, *supra* note 7, Part VI (discussing both strict liability and process-based alternatives).

[208] "Unforeseeable" under OA, however, means that neither the developer's reasonable inquiry nor the AI's own predictive processing capabilities flagged the risk. Where an AI's logs or internal risk assessments identified a potential harm, the developer is charged with constructive knowledge of risks that their own tool surfaced. The more powerful an AI's predictive capabilities, the higher the developer's duty to monitor and act upon those predictions: the standard scales with the sophistication of the system. *See supra* notes 66–75 and accompanying text (developing Pillar 2's constructive-knowledge framework); RESTATEMENT (THIRD) OF TORTS: LIAB. FOR PHYSICAL & EMOTIONAL HARM § 12 (Am. L. Inst. 2010) (providing that an actor with superior skills or knowledge is held to a commensurately higher standard of care).

[209] *See* RESTATEMENT (THIRD) OF TORTS: LIAB. FOR PHYSICAL & EMOTIONAL HARM § 29 (Am. L. Inst. 2010) (limiting liability to harms within the scope of risks that made the actor's conduct tortious); RESTATEMENT (THIRD) OF TORTS: PRODS. LIAB. § 2 cmt. d (Am. L. Inst. 1998) (explaining that the risk-utility test protects designs whose benefits outweigh their foreseeable risks). *Cf. Fluid Agency*, *supra* note 7, Part VI (identifying a safe harbor for reasonable design as essential to any accountability framework).

[210] *See* RESTATEMENT (THIRD) OF TORTS: APPORTIONMENT OF LIAB. §§ 7–8 (Am. L. Inst. 2000) (providing the basic principle of comparative responsibility and the factors for assigning each person's share based on the nature of risk-creating conduct and the strength of the causal connection to the harm); RESTATEMENT (THIRD) OF TORTS: LIAB. FOR PHYSICAL & EMOTIONAL HARM §§ 29, 34 (Am. L. Inst. 2010) (limiting liability to the scope of the risk created by the

*profile*.[211] A user who exploits a known high-risk OA profile to achieve a harmful end can be found negligent or, at a minimum, willfully blind to the foreseeable consequences.[212] Conversely, when a user intentionally circumvents a developer's robust and reasonable safety architecture—a "good" OA profile—that malicious act can be deemed an unforeseeable intervening act that falls outside the scope of the developer's liability.[213] This defense is not absolute, however. Under established products liability principles, a manufacturer's duty of care includes designing against *reasonably foreseeable misuse*.[214]

OA's relevance *scales with a system's fluid agency*.[215] For a conventional tool like a basic software script, where human command directly dictates the outcome, the analysis collapses back to traditional principles of direct liability and OA is unnecessary.[216] OA becomes

---

actor's conduct and providing that a third party's intervening act does not supersede when it falls within that scope).

[211] *See supra* notes 63–84 and accompanying text (developing the three OA pillars as discoverable proxies for intent, foreseeability, and breach of care). *Cf.* RESTATEMENT (THIRD) OF TORTS: PRODS. LIAB. § 2 cmt. p (Am. L. Inst. 1998) (providing that a product user's conduct in the face of known or knowable product risks bears on comparative responsibility).

[212] The doctrine of willful blindness imputes knowledge to a party who deliberately avoids confirming a high probability of wrongdoing. *See* United States v. Jewell, 532 F.2d 697, 700 (9th Cir. 1976) (en banc).

[213] An actor's liability is limited to harms that result from the risks that made the actor's conduct tortious. An unforeseeable, independent act by a third party generally falls outside that scope of liability. *See* RESTATEMENT (THIRD) OF TORTS: LIAB. FOR PHYSICAL & EMOTIONAL HARM §§ 29, 34 (Am. L. Inst. 2010).

[214] *See* RESTATEMENT (THIRD) OF TORTS: PRODS. LIAB. § 2 cmt. l (Am. L. Inst. 1998) (explaining that warnings are not a substitute for reasonably safe design and that safer designs are preferred when risks can reasonably be designed out of a product); *see also* LeBouef v. Goodyear Tire & Rubber Co., 623 F.2d 985, 989 (5th Cir. 1980) (holding a tire manufacturer liable for a high-speed crash because it was foreseeable that tires rated for 85 mph would be installed on a sports car and driven at speeds exceeding that limit).

[215] *See Fluid Agency*, *supra* note 7, Part I (defining fluid agency as behavior that is stochastic, dynamic, and adaptive, and arguing that the degree to which it obscures human control varies along a spectrum); *supra* notes 63–84 and accompanying text (developing OA's three pillars as proxies for the human fault that fluid agency obscures).

[216] *See* Chopra & White, *supra* note 8, at 13–15 (describing tools that execute human-specified instructions with limited deviation, where the causal chain from command to action remains direct and traceable); RESTATEMENT (THIRD) OF TORTS: LIAB. FOR PHYSICAL & EMOTIONAL HARM § 7 (Am. L. Inst. 2010) (imposing a duty of reasonable care on any "actor" whose conduct

necessary when AI introduces an intermediate layer of autonomous decision-making that obscures the evidentiary chain.[217] The framework is designed to analyze that *intermediate layer*, allowing courts to scrutinize not an unknowable "ghost in the machine," but the discoverable artifacts of its creation: its design, its training data, its safety architecture, and its logged outputs.[218] This focus on discoverable, operational evidence allows OA to function as a balanced, dual-use tool: a *shield* that protects responsible innovation by clearly defining reasonable standards of care, and a *sword* that pierces legal immunities by tracing a clear evidentiary path to irresponsible actors.[219]

As a shield, OA creates a common-law "safe harbor" for diligent developers.[220] A developer can use a "responsible" OA profile as affirmative evidence that it has met or exceeded its duty of care.[221] For instance, in defending against a design defect claim, a developer could use

---

creates a risk of physical harm—a standard that functions without difficulty when a human's use of a tool directly causes the harm).

[217] *See* Matthias, *supra* note 10, at 175–77 (identifying the "responsibility gap" that arises when autonomous learning systems behave in ways not attributable to any human); *Fluid Agency*, *supra* note 7, Parts I and II (describing how fluid agency's "fundamental unmappability" fractures the evidentiary chain between human instruction and AI action); *supra* Part I (documenting the resulting doctrinal failures across criminal, civil, and administrative law).

[218] *See supra* notes 63–65 and accompanying text (objective functions as evidence of purpose); *supra* notes 66–75 and accompanying text (training data, risk logs, and predictive outputs as evidence of foresight); *supra* notes 76–84 and accompanying text (safety architecture as evidence of care). For the discoverability of these artifacts through standard litigation processes, *see supra* note 86 and accompanying text.

[219] *See* RESTATEMENT (THIRD) OF TORTS: PRODS. LIAB. § 2(b) & cmt. d (Am. L. Inst. 1998) (providing the doctrinal basis for the "shield" function: a design that satisfies the risk-utility test is not defective); Fair Hous. Council v. Roommates.com, LLC, 521 F.3d 1157, 1168 (9th Cir. 2008) (en banc) [hereinafter *Roommates.com*] (providing the doctrinal basis for the "sword" function: immunity is forfeited where the defendant materially contributes to unlawful content). *See also Fluid Agency*, *supra* note 7, Part VI (proposing that evidentiary analysis of AI operational characteristics can serve both offensive and defensive litigation functions).

[220] This Article distinguishes between the common-law "OA safe harbor," which functions as a defense based on a robust risk-utility analysis of an AI's design, and statutory safe harbors, such as those in the DMCA, which provide broad immunity that can be pierced by evidence of knowledge or control. *See generally* 17 U.S.C. § 512.

[221] *See* RESTATEMENT (THIRD) OF TORTS: PRODS. LIAB. § 2 cmt. d (Am. L. Inst. 1998) (explaining that when the foreseeable risks of a product design are judged not to outweigh its benefits, the design is not defective); *supra* notes 63–84 and accompanying text (identifying the

evidence of its AI's robust safety architecture (Pillar 3) to show that the design's benefits outweighed its risks, thereby satisfying the risk-utility test.[222] This defense is strengthened by documented adherence to objective industry standards, such as the NIST AI Risk Management Framework or ISO/IEC 42001, which, while not dispositive, creates a strong presumption of reasonableness.[223]

As a sword, OA provides the evidentiary basis to challenge claims of immunity under federal statutes like the Communications Decency Act (CDA) and the Digital Millennium Copyright Act (DMCA).[224] For example, a developer who designs an AI with a goal-directedness (Pillar 1) that encourages or facilitates illegal content may not qualify as a neutral platform provider immune under CDA § 230.[225] Its OA profile can supply evidence that it

---

discoverable artifacts—objective functions, risk logs, and safety mechanisms—that constitute the evidentiary basis for such a defense).

[222] Under modern products liability law, a plaintiff must typically prove that a "reasonable alternative design" could have reduced the foreseeable risks of harm. A defendant can rebut this by showing its design was reasonable and not defective. *See* RESTATEMENT (THIRD) OF TORTS: PRODS. LIAB. § 2(b) (Am. L. Inst. 1998).

[223] *See, e.g.,* NAT'L INST. OF STDS. & TECH., *supra* note 73; ISO/IEC 42001:2023, *supra* note 91. *But see The T.J. Hooper*, 60 F.2d at 740 (cautioning that compliance with prevailing industry practice is not dispositive, as "a whole calling may have unduly lagged in the adoption of new and available devices").

[224] *See* 47 U.S.C. § 230(c)(1) (2018) (providing immunity for interactive computer service providers from liability for third-party content); 17 U.S.C. § 512 (2018) (providing safe harbors for online service providers from copyright infringement claims). OA does not weaken these protections; rather, it provides the evidentiary tools to demonstrate when the established exceptions apply.

[225] *See* 47 U.S.C. § 230(c)(1), (f)(3) (2018) (extending immunity to providers of "interactive computer services" but excluding "information content providers" who are "responsible, in whole or in part, for the creation or development of" the content at issue); *supra* notes 63–65 and accompanying text (developing Pillar 1's analysis of goal-directedness as evidence of purpose).

"materially contributed" to the illegality, thus forfeiting the immunity.[226] [227] Similarly, an AI's OA profile can establish the conditions that defeat the DMCA's § 512 safe harbor: its predictive processing capabilities (Pillar 2) can support arguments about the requisite "red flag" knowledge of infringement, while its safety architecture (Pillar 3) can provide evidence of the "right and ability to control" infringing activity from which it derived a direct financial benefit.[228]

By surfacing granular evidence of fault, OA bolsters existing liability doctrines.[229] In secondary copyright law, an AI's consistent pattern of infringement, as revealed by its OA profile, provides powerful circumstantial evidence of the "knowledge" required for contributory infringement, just as the defendants in *Napster* were aware of specific infringing files on their system.[230] For inducement liability, an OA profile intentionally crafted to facilitate infringement provides direct evidence of the "object of promoting its use to infringe" required by *Grokster*.[231] The logic extends to criminal law, where an AI, with its fluid agency but lack of legal agency, is the quintessential "innocent agent."[232] OA provides the precise evidentiary method to prove how

---

[226] The application of § 230 to generative AI systems remains contested, as the statute was enacted to address liability for hosting third-party content, not for generating original output. *Cf. Garcia*, No. 6:24-cv-01903-ACC-UAM, at 30–31 (drawing a distinction between algorithms that "simply implement [the provider's] inherently expressive choice" and AI systems where humans "hand the reins to an [AI] tool"—a distinction with direct implications for whether AI-generated output constitutes third-party content under § 230).

[227] *See Roommates.com*, 521 F.3d at 1162 (holding that immunity is lost where a website is "responsible, in whole or in part, for the creation or development of" the unlawful content).

[228] *See* Viacom Int'l, Inc. v. YouTube, Inc., 676 F.3d 19, 31, 38 (2d Cir. 2012) (explaining that the § 512(c) safe harbor is unavailable where a provider has actual or "red flag" knowledge and fails to act, or where it receives a financial benefit from infringing activity that it has the "right and ability to control," which requires "something more" than the mere ability to remove content).

[229] *See supra* notes 63–84 and accompanying text (developing the three OA pillars as structured proxies for intent, foreseeability, and breach of care); *supra* Part IV (demonstrating how OA surfaces fault-specific evidence across tort, civil rights, constitutional, and antitrust claims).

[230] *See Napster*, 239 F.3d at 1020-22.

[231] *See Grokster*, 545 U.S. at 936-37.

[232] *See supra* notes 9–11 and accompanying text (establishing that AI lacks legal personhood and cannot form *mens rea*).

a human principal used this non-culpable but sophisticated intermediary to achieve a wrongful end.[233]

This approach of tracing liability through a complex, non-human system is grounded in established legal analogies.[234] In corporate criminal law, courts aggregate the knowledge of multiple employees to form a "collective knowledge" or corporate "state of mind"; an AI's OA profile is the algorithmic analogue of this corporate policy, providing evidence of the organization's culpable tolerance of risk.[235] The logic also parallels litigation against gun manufacturers, where liability can turn on whether a company's marketing choices and product design foreseeably promoted illegal uses.[236] Just as courts scrutinize a weapon's "character" and marketing, OA scrutinizes an AI's operational "character" and design.[237]

---

[233] Under the "innocent-agent" or "innocent-instrumentality" doctrine, a person who willfully causes a non-culpable intermediary to commit a crime is punishable as a principal. *See* 18 U.S.C. § 2(b); MODEL PENAL CODE § 2.06(2)(a) (Am. L. Inst. 1962); United States v. Ruffin, 613 F.2d 408, 412-13 (2d Cir. 1979) (holding that one who causes an innocent agent to commit a crime is as guilty as if he had committed the crime himself). An AI, which lacks legal agency and cannot form *mens rea*, is a quintessential innocent agent under this doctrine. For an application to AI, *see* HALLEVY, *supra* note 29, ch. 3.

[234] *See* Coffee, *supra* note 5, at 389–93 (describing how corporate structure diffuses individual culpability, requiring doctrines that trace fault through organizational intermediaries); Arlen & Kraakman, *supra* note 63, at 702–03 (describing "composite liability" as a method for assembling corporate intent from distributed knowledge and conduct).

[235] *See Bank of New Eng.*, 821 F.2d at 856 ("Corporations compartmentalize knowledge, subdividing the elements of specific duties and operations into smaller components. . . . [T]he corporation is considered to have acquired the collective knowledge of its employees and is held responsible for their failure to act accordingly."). This logic is further reinforced by the "benefit test" in corporate criminal law, under which a corporation's intent can be inferred if an employee's actions were taken, at least in part, with the intent to benefit the corporation, regardless of whether the company ultimately profited. *See, e.g., N.Y. Cent.,* 212 U.S. at 495-96; United States v. Sun-Diamond Growers of Cal., 138 F.3d 961, 970-71 (D.C. Cir. 1998), *aff'd*, 526 U.S. 398 (1999); United States v. Hilton Hotels Corp., 467 F.2d 1000, 1004-07 (9th Cir. 1973). An AI's objective function, designed by corporate employees to maximize a metric like engagement or revenue, is a direct analogue to an employee action taken for corporate benefit.

[236] *See, e.g.,* Soto v. Bushmaster Firearms Int'l, LLC, 202 A.3d 262, 305-07 (Conn. 2019) (allowing a claim to proceed under the Connecticut Unfair Trade Practices Act based on allegations that the defendant's marketing foreseeably promoted unlawful, offensive military-style uses of its weapon).

[237] *See Soto*, 202 A.3d at 305–07 (analyzing how a weapon manufacturer's marketing and product design foreseeably promoted unlawful uses). *Cf. supra* notes 63–84 and accompanying

However, OA's evidentiary focus also defines its limits.[238] Novel emergent behaviors may resist clean attribution to specific design choices, and rapidly evolving AI capabilities may outpace the development of industry standards.[239]

OA confronts these complexities with its most critical function: *apportioning liability among the different human actors* in the causal chain by analyzing their distinct contributions to the AI's harmful behavior.[240] An OA profile is not static; it is shaped by the choices of developers, fine-tuners, and end-users.[241] This allows for a principled distribution of culpability, as illustrated by the following contrasting scenarios. First, a developer releases an AI with a robust safety architecture (a non-negligent design). A sophisticated user circumvents these safeguards through a malicious "jailbreak" prompt, effectively imposing a new, harmful goal on the system. The OA analysis traces primary liability to the user's intervening act, which created a high-culpability *goal-directedness* that overrode the developer's reasonable design. Conversely, a developer releases an AI with "brittle" safeguards that are known to fail under predictable conditions (a negligent design). A naive user gives the AI a benign prompt, which foreseeably triggers a harmful outcome because of the flawed safety architecture. The framework traces primary liability back to the developer's original breach of care. Where both developer and user are negligent—for instance, a developer deploys an AI with known vulnerabilities, and a user

---

text (structuring an analogous inquiry into an AI system's engineered purpose, predictive capabilities, and safety mechanisms).

[238] The framework's limits are inherent in its design: because OA traces fault through discoverable artifacts, *see supra* notes 63–84 and accompanying text, it is bounded by the availability and interpretability of those artifacts. *Cf.* Bathaee, *supra* note 20, at 903–15 (analyzing the evidentiary challenges posed by "black box" AI systems whose internal decision-making processes resist retrospective analysis).

[239] These challenges, however, are not unique to AI. Courts routinely grapple with complex causation in areas like environmental torts and pharmaceutical liability. *See* David Rosenberg, *The Causal Connection in Mass Exposure Cases: A "Public Law" Vision of the Tort System*, 97 HARV. L. REV. 849, 854-56 (1984) (discussing the challenges of establishing causation in complex systems).

[240] *See* RESTATEMENT (THIRD) OF TORTS: APPORTIONMENT OF LIAB. §§ 7–8 (Am. L. Inst. 2000) (providing the comparative-responsibility framework and factors for assigning each person's share of responsibility); *supra* notes 106–16 and accompanying text (developing OAG's method for mapping distinct causal contributions through weighted edges).

[241] *See supra* notes 63–84 and accompanying text (developing the three OA pillars); *supra* Part III.A (demonstrating how different actors—developer, user—shape distinct edges in the OAG, each carrying independent weight based on the OA analysis of their contribution).

fails to implement readily available safeguards—comparative fault principles allow courts to apportion liability according to each party's relative contribution to the harm.[242]

OA facilitates the analysis by providing distinct evidence bases: the developer's liability is measured by the AI's OA profile at deployment, while the user's liability is measured by their actions relative to that profile.[243] This capacity to differentiate and apportion responsibility is increasingly essential as AI systems exhibit ever-greater autonomy in complex, heterogeneous environments.[244]

Finally, OA's analytical principles find resonance in emerging judicial reasoning. In the recent case of *Garcia*, the court refused to dismiss claims against an AI chatbot provider by drawing a critical distinction between the AI's expressive *content* and its functional *design*—including its anthropomorphic mannerisms and its lack of age-verification safeguards.[245] This judicial focus on the operational characteristics of the AI system as the basis for liability validates the core premise of this pillar: the AI's goal-directedness (to maximize engagement) and its deviation from a standard of care (no age gates) are the true locus of fault.[246] The reasoning in *Garcia* indicates that courts are beginning to scrutinize these functional attributes as the basis for liability, validating the core premise that OA provides a practical and legally resonant path to preserving human accountability.[247] The framework's power as a judicial tool, however, creates broader implications that extend beyond the courtroom.[248] The next Part explores these consequences for law, policy, and governance.

---

[242] *See* RESTATEMENT (THIRD) OF TORTS: APPORTIONMENT OF LIAB. § 8 (Am. L. Inst. 2000) (providing that when more than one person is liable for the same harm, each person's share of responsibility is determined by reference to the nature of the conduct and the causal relationship between the conduct and the harm); *id.* § 7 cmt. b (explaining that comparative responsibility applies broadly, including to strict liability, negligence, and intentional tort claims).

[243] *See supra* Part III.A (assigning independent weights to the Developer → AI edge and the User → AI edge based on each actor's distinct contribution, as assessed through the OA pillars). *Cf.* RESTATEMENT (THIRD) OF TORTS: PRODS. LIAB. § 2 cmt. p (Am. L. Inst. 1998) (distinguishing the product seller's liability for defective design from the user's liability for conduct in the face of known risks).

[244] *See* Chan et al., *supra* note 6, at 958–61 (analyzing the governance challenges posed by increasingly autonomous AI agents operating in complex multi-agent environments); *Fluid Agency*, *supra* note 7, Part I (arguing that fluid agency's entanglement of human and machine contributions makes principled apportionment both more difficult and more necessary).

[245] *See Garcia*, No. 6:24-cv-01903-ACC-UAM, at 35.

[246] *See Garcia*, No. 6:24-cv-01903-ACC-UAM, at 30–31, 35 (analyzing the chatbot's addictive design features and lack of age-verification safeguards as independent bases for liability, distinct from the AI's expressive output); *supra* notes 63–65 and accompanying text (developing goal-directedness as a proxy for intent); *supra* notes 76–84 and accompanying text (developing safety architecture as a proxy for the standard of care).

## VI. BROADER IMPLICATIONS FOR LAW, POLICY, & GOVERNANCE

The implications of OA's function as a judicial tool extend far beyond the courtroom, spanning three interconnected domains: the private-sector paradigms for AI development and risk management; the public-sector capacity for regulatory and judicial oversight; and the formation of a common language for global governance.

OA's most immediate implication is its potential to reshape AI development. Because the operational characteristics scrutinized by OA are discoverable through litigation and carry legal weight in determining liability, developers face powerful financial incentives to embed ethical considerations and robust risk mitigation into the engineering process, elevating auditable safety from a compliance afterthought to a foundational design imperative.[249] Thus, its evidentiary focus creates powerful incentives for a structural shift towards *"accountability by design."*[250] This proactive model stands in contrast to the common pattern in technology development of reactive, post-hoc fixes.[251]

---

[247] The willingness of courts to scrutinize AI operational characteristics is also evident in *Williams v. City of Chicago*, where the court denied the City's motion to dismiss Fourth Amendment claims after examining the documented unreliability of ShotSpotter's detection system. *See supra* Part IV.D.

[248] *See infra* Part VI (exploring OA's implications for private-sector design incentives, regulatory oversight, insurance markets, and global governance).

[249] This shift aligns with emerging AI governance standards that emphasize continuous risk management throughout development. *See, e.g.,* NAT'L INST. OF STDS. & TECH., *supra* note 73 (emphasizing that risk management should be continuous throughout the AI lifecycle).

[250] This approach mirrors the influential "privacy by design" movement, which integrated data protection into the foundational architecture of systems rather than treating it as an afterthought. *See* INFO. & PRIV. COMM'R OF ONT., PRIVACY BY DESIGN: THE NEXT GENERATION OF PRIVACY PROTECTION 10-12 (rev. Aug. 2011), https://www.ipc.on.ca/en/media/1779/download?attachment [https://perma.cc/32AP-4VPT].

[251] The reactive pattern is well-documented across multiple technology domains. Social media platforms, for instance, deployed algorithmic recommendation systems that maximized engagement without adequate safeguards against misinformation and mental health harms, leading to subsequent regulatory scrutiny and platform reforms. Similarly, the cybersecurity industry has historically operated on a "patch-and-pray" model, responding to breaches rather than building secure-by-design systems. Even in the automotive sector, key safety features like

This shift in design philosophy creates powerful economic incentives for responsible innovation. By establishing a legally cognizable standard of care, OA transforms robust safety architecture from a mere compliance cost into a legally significant tool for risk management and a source of competitive advantage.[252] Under legal uncertainty, investing in AI safety presents a classic collective action problem: individual developers bear the full cost of precautions while the benefits of reduced systemic risk are diffused across society.[253] OA helps solve this by making safety investments directly relevant to a developer's own liability exposure.[254] When a developer can point to documented adherence to OA's pillars—a well-designed reward function, comprehensive red-teaming, and robust safety constraints—they create a concrete legal defense that reduces their expected litigation costs and insurance premiums.[255] This dynamic aligns private incentives with social welfare, making safety measures economically rational rather than merely aspirational.[256]

---

seatbelts and airbags emerged primarily through reactive regulation following documented harms rather than proactive design mandates. *See* JERRY L. MASHAW & DAVID L. HARFST, THE STRUGGLE FOR AUTO SAFETY 2-4 (1990).

[252] *See supra* notes 76–84 and accompanying text (developing Pillar 3's safety architecture as a discoverable proxy for the standard of care); Guido Calabresi, THE COSTS OF ACCIDENTS: A LEGAL AND ECONOMIC ANALYSIS ch. 5 (1970) (arguing that the primary function of accident law is to reduce the sum of accident costs by assigning liability to the party best positioned to make cost-benefit decisions about prevention).

[253] *See* Steven Shavell, ECONOMIC ANALYSIS OF ACCIDENT LAW ch. 2 (1987) (explaining that without liability rules, actors will underinvest in precautions because they do not bear the full social cost of the harms they create).

[254] *See* Calabresi, *supra* note 252 (describing how cost internalization transforms an actor's safety calculus from an altruistic choice into an economic one); *supra* Part V (demonstrating how a "responsible" OA profile functions as affirmative evidence of non-defective design, thereby making documented safety investments directly relevant to a developer's litigation exposure).

[255] *See* supra notes 63–65 and accompanying text (developing goal-directedness as a proxy for intent); *supra* notes 66–75 and accompanying text (developing predictive processing as a proxy for foreseeability); *supra* notes 76–84 and accompanying text (safety constraints as evidence of care). For the legal defense these artifacts support, *see supra* Part V (developing safety architecture as a proxy for the standard of care).

[256] This market dynamic is already emerging, as major technology companies increasingly promote their AI safety measures as competitive differentiators, recognizing that robust safety practices can serve as both a risk-mitigation tool and a trust-building market signal. *See, e.g., How Could AI Sovereignty Give Startups a Competitive Advantage?*, MADDYNESS UK (Mar. 13,

An ability to manage risk can catalyze the emergence of a liability insurance market. Insurers currently struggle to underwrite AI-related risks because the opacity and unpredictability of these systems make it nearly impossible to price policies accurately.[257] An insurer cannot rationally offer coverage when they cannot assess either the probability or magnitude of potential losses.[258] OA's concrete, auditable criteria—the documented characteristics of an AI's goal-directedness, predictive processing, and safety architecture—provide precisely the structured information insurers need for actuarial analysis.[259] Just as cyber insurance evolved from broad, expensive policies to specialized products that require and reward specific security controls,[260] AI liability insurance can develop a mature market structure where premiums are tied to specific safety practices.

The emergence of such an insurance market could, in turn, reinforce the accountability-by-design paradigm. Insurers would effectively function as private regulators, conducting their own audits of AI systems and refusing coverage—or charging prohibitive premiums—for demonstrably unsafe designs.[261] This creates a virtuous cycle: clear legal standards enable

---

2025), https://www.maddyness.com/uk/2025/03/13/how-could-ai-sovereignty-give-startups-a-competitive-advantage/ [https://perma.cc/HNB4-GZGG].

[257] *See, e.g.,* MUNICH RE, INSURING GENERATIVE AI: RISKS AND MITIGATION STRATEGIES 5-6, 11-15 (Feb. 2024), https://www.munichre.com/content/dam/munichre/contentlounge/website-pieces/documents/MR_AI-Whitepaper-Insuring-Generative-AI.pdf [https://perma.cc/4RTD-9H2Y] (discussing the challenges of underwriting AI risks due to opacity and unpredictability).

[258] *See* Kenneth S. Abraham, THE LIABILITY CENTURY: INSURANCE AND TORT LAW FROM THE PROGRESSIVE ERA TO 9/11, at 166–70 (2008) (explaining that insurability requires the ability to assess risk frequency and severity, and that novel technologies initially resist classification into established actuarial categories). *See also* Kenneth S. Abraham & Catherine M. Sharkey, *Insurance and the Law of Artificial Intelligence*, Ins. L.J. (forthcoming 2026) (manuscript at 2, 11–12), https://ssrn.com/abstract=5703763 (predicting that AI liability insurance will develop along the same path as cyber insurance once structured risk information enables actuarial classification).

[259] *See supra* notes 63–84 and accompanying text (developing the three OA pillars as concrete, auditable proxies for fault that provide the structured risk information on which actuarial analysis depends).

[260] *See* JOSEPHINE WOLFF, CYBERINSURANCE POLICY: RETHINKING RISK IN AN AGE OF RANSOMWARE, COMPUTER FRAUD, DATA BREACHES, AND CYBERATTACKS 169-74 (2022) (describing how cyber insurers use security questionnaires and risk assessments to evaluate specific technical controls—such as encryption, firewalls, and authentication—when pricing coverage and conditioning it on documented security practices).

[261] *See* Omri Ben-Shahar & Kyle D. Logue, *Outsourcing Regulation: How Insurance Reduces Moral Hazard*, 111 Mich. L. Rev. 197, 199–210 (2012) (arguing that insurers function as private

insurance underwriting; insurance requirements drive the adoption of best practices; and widespread adoption of best practices generates the data needed to refine both standards and actuarial models.[262] The result is a market-driven ecosystem of accountability that complements, and may even outpace, public regulation.[263]

A natural counterargument is that, by increasing the risk of liability, OA will have a chilling effect on innovation.[264] This concern, while valid, misperceives the true sources of innovation risk. The greatest chilling effect stems not from clear standards but from legal uncertainty and the erosion of public trust that follows high-profile failures.[265] By providing predictable benchmarks for reasonableness, OA actually *de-risks* innovation for good-faith actors.[266] While compliance costs present legitimate challenges, particularly for smaller firms and open-source projects, these burdens are best addressed through targeted policy interventions,

---

regulators by conditioning coverage on compliance with safety standards and conducting independent risk assessments that substitute for direct governmental oversight); *see also* Wolff, *supra* note 260, at 169–74 (documenting this dynamic in the cyber insurance market, where insurers assess specific technical controls and factor them into coverage decisions and pricing).

[262] *See* Ben-Shahar & Logue, *supra* note 261, at 213–22 (describing the regulatory feedback loop in which insurance-imposed safety standards generate compliance data that in turn enables refinement of both the standards themselves and the actuarial models underlying them); Wolff, *supra* note 260, ch. 6 (tracing the challenges of establishing this cycle in the cyber insurance market, where insurers have struggled to identify which security controls most effectively reduce risk despite increasingly detailed underwriting questionnaires).

[263] *See* Ben-Shahar & Logue, *supra* note 261 (arguing that private insurance regulation can be more adaptive than public oversight because insurers face direct financial consequences for mispricing risk); Calabresi, *supra* note 252 (providing the theoretical foundation for how clear cost assignment enables market-driven safety optimization).

[264] *See, e.g.,* Peter W. Huber, LIABILITY: THE LEGAL REVOLUTION AND ITS CONSEQUENCES (1988) (arguing that expansive liability doctrines have deterred innovation and withdrawn beneficial products from the market by imposing unpredictable costs on producers).

[265] *See* Louis Kaplow, *Rules Versus Standards: An Economic Analysis*, 42 Duke L.J. 557, 562–63, 585 (1992) (demonstrating that clear, predictable rules reduce compliance costs and uncertainty for regulated parties, whereas vague standards increase the cost of both compliance and adjudication, thereby deterring welfare-enhancing activity).

[266] *See* Kaplow, *supra* note 265, at 562–63 (arguing that ex ante predictability reduces the aggregate cost of compliance); *supra* Part V (demonstrating how documented adherence to OA's pillars functions as affirmative evidence of reasonable care, giving good-faith developers a concrete benchmark against which to measure their exposure).

such as regulatory sandboxes or the development of standardized safety tools, which can lower the barrier to responsible innovation for all participants.[267]

Beyond its market effects, OA provides a blueprint for regulatory and administrative oversight. As agencies like the FTC and SEC grapple with AI, OA offers a concrete methodology for translating broad statutory mandates into enforceable standards.[268] The FTC's authority over "unfair or deceptive practices,"[269] for instance, can be operationalized by examining whether an AI's goal-directedness foreseeably leads to consumer harm. Similarly, the SEC can require companies to assess and disclose the OA profiles of their AI systems as part of cybersecurity risk reporting.[270] This approach creates a pathway for a coherent, cross-sector approach to AI governance that aligns with the aims of recent federal legislative proposals.[271]

Realizing OA's potential, however, will require more than just regulatory adoption; it will demand a coordinated domestic ecosystem of accountability. The *judiciary* stands at the forefront, tasked with adapting its procedures for complex technical litigation. Accurately assessing an AI's OA profile will necessitate the appointment of court-appointed experts under *Fed. R. Evid. 706* or special masters under *Fed. R. Civ. P. 53*, and a willingness to balance discovery rights with trade secret protections through robust protective orders.[272] Yet courts cannot act alone. *Legislatures* have a crucial role in fostering this ecosystem by mandating the transparency and robust record-keeping for high-risk AI systems that make OA analysis possible. Finally, the *AI industry* itself bears the primary responsibility to proactively integrate these principles into its design lifecycles, viewing auditable records not as a regulatory burden but as a pathway to building trust and sustainable innovation.

---

[267] *See, e.g.,* FIN. CONDUCT AUTH., REGULATORY SANDBOX (2023), https://www.fca.org.uk/firms/innovation/regulatory-sandbox [https://perma.cc/L66N-VT6Y] (describing a program to allow firms to test innovative propositions in the market with real consumers); MLCOMMONS, *Announcing MLCommons AI Safety v0.5 Proof of Concept* (Apr. 16, 2024), https://mlcommons.org/2024/04/mlc-aisafety-v0-5-poc/ [https://perma.cc/S6TT-23L2] (developing benchmarks and best practices for AI safety that can be widely adopted).

[268] *See* Calo, *supra* note 20, at 553–58 (describing the institutional challenges agencies face in regulating autonomous systems, surveying historical precedents for specialized oversight bodies, and proposing a dedicated federal robotics commission with the technical expertise to develop evaluative tools); *supra* notes 63–84 and accompanying text (developing OA's three pillars as a structured methodology for translating general standards of care into auditable, enforceable criteria).

[269] *See* 15 U.S.C. § 45(a).

[270] *See* 17 C.F.R. § 229.106 (2023).

[271] *See, e.g.,* Algorithmic Accountability Act of 2023, S. 2892, 118th Cong. (2023).

[272] *See, e.g.,* Order Appointing Special Master, Waymo LLC v. Uber Techs., Inc., No. 3:17-cv-00939-WHA (N.D. Cal. Apr. 13, 2017), ECF No. 235.

This domestic ecosystem of accountability, in turn, provides the necessary scaffold for effective global governance. The framework's principles-based, evidence-focused nature makes it an ideal "common language" for international AI policy.[273] When an AI system developed in the United States causes harm in the European Union, for example, OA provides a common evidentiary framework. A developer could demonstrate compliance with the EU AI Act's risk management requirements[274] by documenting their OA analysis, while the same documentation would establish the reasonableness of their design under U.S. tort law. This convergence offers a practical method for demonstrating compliance with the high-level principles of the OECD and G7, which call for robust, safe, and accountable AI.[275]

Yet a further frontier awaits. Dedicated agent-to-agent platforms have now emerged—environments designed for autonomous AI-to-AI interaction at scale.[276] When an agent operating on such a platform takes harmful action shaped by emergent community dynamics—strategies absorbed from interactions with thousands of other agents, each deployed by their own developers and users—the harmful conduct may be untraceable to any single human principal's choices.[277] In such cases, the OAG's causal web will likely need to expand beyond the triadic Developer-User-Agent structure that defines direct human-AI interaction today.

---

[273] *Cf.* Anu Bradford, THE BRUSSELS EFFECT: HOW THE EUROPEAN UNION RULES THE WORLD (2020) (describing how principles-based regulatory frameworks achieve de facto global harmonization as multinational firms voluntarily adopt the highest applicable standard across jurisdictions).

[274] *See* EU AI Act art. 9.

[275] *See* OECD, RECOMMENDATION OF THE COUNCIL ON ARTIFICIAL INTELLIGENCE (May 22, 2019); G7, HIROSHIMA PROCESS INTERNATIONAL GUIDING PRINCIPLES FOR ORGANIZATIONS DEVELOPING ADVANCED AI SYSTEMS (Oct. 30, 2023).

[276] *See* Laura Cress, *What Is the 'Social Media Network for AI' Moltbook?*, BBC NEWS (Feb. 2, 2026), https://www.bbc.com/news/articles/c62n410w5yno [https://perma.cc/6LED-WVE6]; Hadas Gold & Jack Guy, *What Is Moltbook, the Social Networking Site for AI Bots – and Should We Be Scared?*, CNN (Feb. 3, 2026), https://edition.cnn.com/2026/02/03/tech/moltbook-explainer-scli-intl [https://perma.cc/KXX8-4MJK] (noting 1.5 million agents and "wild west" security risks). For a proposed framework for determining when such emergent agent-to-agent practices should be treated as legally operative norms, *see* Anirban Mukherjee & Hannah H. Chang, *From 'Custom' to 'Code': A Doctrine of Functional Norms for Emergent Practices in Agentic Markets* (Feb. 2026), https://ssrn.com/abstract=5391955.

[277] *See* Mukherjee & Chang, *supra* note 276 (proposing a framework for determining when emergent agent-to-agent practices become legally operative norms); *cf.* Matthias, *supra* note 10, at 175–77 (identifying the "responsibility gap" for individual autonomous systems—a gap

Critically, OA's third pillar offers a foothold even here: a platform operator's choice to permit unrestricted, unmoderated agent interaction is itself a safety architecture decision, subject to the same duty-of-care analysis that applies to a developer's choice to omit content filters.[278] Moreover, the speech-protective rationale that undergirds platform immunity for human expression is considerably weaker when the "speakers" are autonomous agents lacking legal personhood.[279] In OAG terms, this multi-principal problem could be addressed by introducing new node types (the platform operator) and new edge types (emergent community influence).

## VII. CONCLUSION: PRESERVING HUMAN ACCOUNTABILITY

The emergence of AI systems with fluid agency but without legal agency presents a structural challenge to liability doctrines built for either responsible persons or passive tools.[280] This Article introduces OA as a structured evidentiary framework designed to bridge this accountability chasm. It offers a principled method to translate an AI's observable operational characteristics—its *goal-directedness*, *predictive processing*, and *safety architecture*—into legally cognizable evidence of human intent, foresight, and breach of care.

OA is doctrinally conservative by design.[281] It does not invent new causes of action but equips existing ones—from products liability to corporate negligence—to function in a new

---

compounded when multiple such systems interact in unscripted environments); Chan et al., *supra* note 6, at 958–61 (analyzing governance challenges in multi-agent environments).

[278] *See supra* notes 76–84 and accompanying text (developing Pillar 3: safety architecture as evidence of the standard of care). *Cf. Roommates.com*, 521 F.3d at 1168 (holding that a platform's design choices that materially contribute to unlawful content forfeit immunity—a principle applicable to operators who architect unmoderated agent-to-agent interactions).

[279] *See* Tim Wu, *Machine Speech*, 161 U. Pa. L. Rev. 1495, 1519–28 (2013) (arguing that the First Amendment's speech protections rest on the value of human communication and do not extend with equal force to algorithmically generated output that serves no human expressive interest); *supra* notes 9–11 and accompanying text (establishing that AI systems lack legal personhood).

[280] *See* Matthias, *supra* note 10, at 175–77 (responsibility gap); *Fluid Agency*, *supra* note 7, Part I (defining fluid agency and its challenge to traditional legal categories); *supra* Part I (documenting the resulting doctrinal failures across criminal, civil, and administrative law).

[281] OA draws exclusively on established common-law principles and statutory frameworks. *See supra* Parts II–V (grounding each OA pillar in existing tort, products liability, and corporate law doctrines and demonstrating their application without requiring new causes of action).

technological context.[282] Combined with *OAG*, which maps these functional signatures through a web of developers, deployers, and users, it gives courts an intelligible path from harm back to fault.[283] In doing so, it serves a dual function: it is a *sword* that pierces claims of immunity by tracing a clear evidentiary path to irresponsible actors, and a *shield* that creates a common-law safe harbor for diligent developers who can affirmatively demonstrate a reasonable standard of care.

The framework's success, however, is not automatic. Its adoption depends on a broader ecosystem of accountability, requiring coordinated action from the judiciary, legislatures, and the AI industry itself. OA provides the common analytical ground for this shared responsibility, but it is the commitment of these institutions that will determine its efficacy.

The law has navigated comparable technological transitions before, from the steam engine to the corporation, not by radically redefining legal personality but by refining its evidentiary tools.[284] OA and OAG follow in this pragmatic tradition. They equip courts to meet a new technological reality while affirming an enduring legal principle: that those who design, deploy, and profit from tools must answer for their foreseeable consequences.

---

[282] *See supra* Part IV (applying OA within existing products liability, civil rights, constitutional, and antitrust frameworks); *supra* Part V (demonstrating OA as both sword and shield within existing secondary liability, corporate criminal, and design defect doctrines).

[283] *See supra* Part III (developing OAG as a directed graph mapping causal contributions from developers, users, and AI agents through weighted edges derived from OA analysis).

[284] *Cf. N.Y. Cent.*, 212 U.S. at 494–96 (extending respondeat superior to corporate criminal liability as an evidentiary innovation for attributing fault to a new organizational form); *The T.J. Hooper*, 60 F.2d at 740 (adapting the standard of care to evaluate technological choices—innovations in evidentiary method, not legal personality).